\documentclass[12pt,twoside]{article}
\usepackage{psfig,amsfonts}
\newcommand{\VolumeHeader}{}
\newcommand{\VolumeSerial}{}

\newcommand{\ActivityName}{ {\normalsize {\it 
Spring School on Superstring Theory and Related Topics
}}}
\newcommand{\ActivityDate}{ {\normalsize {\it
Trieste, 31 March-8 April 2003
}}}

\newcommand{\be}{\begin{equation}}
\newcommand{\ee}{\end{equation}}
\newcommand{\bea}{\begin{eqnarray}}
\newcommand{\eea}{\end{eqnarray}}

\newcommand{\LectureHeader}{Light-Cone String Field Theory in a Plane
Wave Background}

\begin{document}
\pagestyle{myheadings}
\markboth{\LectureHeader}{\VolumeHeader}
\markright{\VolumeHeader}

%%%%%%%%%%%%%%%%%%%%%%%%%%%%%%%%%%%%%%%%%%%%%%%%%%%%%%%%%%%%%%%%%%%%%%%%%%%
%%%            Title page starts here                                     %
%%%%%%%%%%%%%%%%%%%%%%%%%%%%%%%%%%%%%%%%%%%%%%%%%%%%%%%%%%%%%%%%%%%%%%%%%%%

\begin{titlepage}

%%% YOUR CHANGES BELOW THIS LINE

\title{Light-Cone String Field Theory in a Plane
Wave}

\author{Marcus Spradlin
%$^\dagger$
and Anastasia Volovich
%$^\ddagger$
\\[1cm]
{\normalsize
%{\it $^\dagger$ Princeton University, Princeton NJ, USA.}
}
%\\
{\normalsize
{\it
%$^\ddagger$
Kavli Institute for Theoretical Physics, Santa Barbara
CA, USA.}}
\\[10cm]
%%% FOR FURTHER AUTHORS SEE WHAT IT IS WRITTEN IN THE ABSTRACT 
%%% DO NOT CHANGE THE FOLLOWING LINES
{\normalsize {\it Lectures given by M.S. at the: }}
\\
\ActivityName 
\\
\ActivityDate 
\\[1cm]
{\small \VolumeSerial} 
}
\date{}
\maketitle
\thispagestyle{empty}
\end{titlepage}

\baselineskip=14pt
\newpage
\thispagestyle{empty}

%%%%%%%%%%%%%%%%%%%%%%%%%%%%%%%%%%%%%%%%%%%%%%%%%%%%%%%%%%%%%%%%%%%%%%%%%%%
%%%            Abstract page starts here                                  %
%%%%%%%%%%%%%%%%%%%%%%%%%%%%%%%%%%%%%%%%%%%%%%%%%%%%%%%%%%%%%%%%%%%%%%%%%%%

{~}
\bigskip
\bigskip
\bigskip
\bigskip
\bigskip
\bigskip
\bigskip
\bigskip
\bigskip
\bigskip
\bigskip
\bigskip

\begin{abstract}

%%% YOUR CHANGES BELOW THIS LINE

These lecture notes present an elementary
introduction to
light-cone string field theory, with an emphasis on its
application to the study of string
interactions in the plane wave limit of AdS/CFT.
We summarize recent results and conclude with a list of open
questions.

\end{abstract}

%\vspace{6cm}

%{\it Keywords:} [...]

%{\it PACS numbers:}
%[...]

%%%%%%%%%%%%%%%%%%%%%%%%%%%%%%%%%%%%%%%%%%%%%%%%%%%%%%%%%%%%%%%%%%%%%%%%%%%
%%%       Automatic TOC and your Text starts here                         %
%%%%%%%%%%%%%%%%%%%%%%%%%%%%%%%%%%%%%%%%%%%%%%%%%%%%%%%%%%%%%%%%%%%%%%%%%%%

\newpage
\thispagestyle{empty}
\tableofcontents

\newpage
\setcounter{page}{1}

\section{Introduction}

\subsection{Light-Cone String Field Theory}

These lectures are primarily about light-cone string field theory,
which is an ancient subject (by modern standards) whose origins lie
in the days of `dual resonance models' even before string theory was
studied as a theory of quantum gravity
\cite{Mandelstam:jk,Mandelstam:hk,Cremmer:1974jq,Cremmer:1974ej,
Green:1982tc,Green:hw,Green:fu}.
Light-cone string field theory is nothing more than the study
of string theory (especially string interactions) via Hamiltonian
quantization in light-cone gauge.  Let us immediately illustrate this point.

Textbooks on string theory typically begin by considering
the Polyakov action for a free string,
\begin{equation}
\label{eq:flatspace}
S = {1 \over 4 \pi \alpha'} \int_{\Sigma} \sqrt{\gamma} \gamma^{\alpha \beta}
\partial_\alpha X^\mu {\partial}_\beta X_\mu
+ \mbox{fermions} + \mbox{ghosts},
\end{equation}
where $\Sigma$ is the $1+1$-dimensional string worldsheet
(we consider only closed strings in these lectures), $\gamma$
is the metric on $\Sigma$, and $X^\mu$ is the embedding of
the string worldsheet into spacetime.  
Quantizing this theory is simplest in light-cone gauge, where the
ghosts are not needed, and one finds the light-cone Hamiltonian
\begin{equation}
\label{eq:hflat}
H = 2 p^- = {1 \over p^+} \left[
{1 \over 2}
p^i p^i + {1 \over \alpha'}
\sum_{n=1}^\infty(\alpha_{-n}^i\alpha_n^i + \tilde{\alpha}_{-n}^i
\tilde{\alpha}_n^i) + \mbox{fermions}
\right],
\end{equation}
where $i$ labels the directions transverse to the light-cone.
At this stage most textbooks abandon light-cone gauge
in favor of more powerful and mathematically beautiful covariant
techniques.

However, it is also possible to continue by second-quantizing
(or third-quantizing, depending on how you count) the
Hamiltonian~(\ref{eq:hflat}).
To do this we introduce a multi-string Hilbert space, with
operators acting to create or annihilate entire strings
(not to be confused with the operators $\alpha_n^i$ and
$\tilde{\alpha}_n^i$, which
create or annihilate an oscillation of frequency $n$ on a given
string).    All of the details will be presented in Lecture~2, where
our ultimate goal will be to write down a relatively simple
interaction term which (at least for the bosonic string) is able
to reproduce (in principle) all possible string scattering amplitudes!

Light-cone string field theory 
obscures many important properties of string theory which are manifest
in a covariant treatment.  
Nevertheless, the subject has recently enjoyed a remarkable
renaissance following \cite{Spradlin:2002ar} because it is
well-suited for studying string interactions in a maximally symmetric
plane wave background,
where
covariant techniques are more complicated than they are in flat space.

\subsection{The Plane Wave Limit of AdS/CFT}

One of the most exciting developments in string theory
has been the discovery of the AdS/CFT correspondence 
\cite{Maldacena:1997re}
(see
\cite{Aharony:1999ti} for a review).  The best understood
example of this correspondence relates type IIB string theory
on $AdS_5 \times S^5$ to SU($N$) Yang-Mills theory with
${\cal{N}} = 4$ supersymmetry.  Although we have learned a
tremendous amount about gauge theory, quantum gravity, and
the holographic principle from AdS/CFT, the full promise of the
duality has unfortunately not yet been realized.  The reason is
simple: string theory on $AdS_5 \times S^5$ is hard!

In contrast to~(\ref{eq:flatspace}),
the Green-Schwarz superstring on $AdS_5 \times S^5$ is
nontrivial.  It can be
regarded as a coset sigma model on $PSU(2,2|4)/SO(4,1) \times SO(5)$
with an additional fermionic Wess-Zumino term and a
fermionic $\kappa$-symmetry \cite{Metsaev:1998it,Kallosh:1998zx,Roiban:2000yy}.
Despite some intriguing recent progress \cite{Bena:2003wd,Dolan:2003uh},
this theory remains intractable and
we still do not know the free string spectrum (i.e., the
analogue of~(\ref{eq:hflat})) in $AdS_5 \times S^5$.
Because of this difficulty, most applications of the AdS/CFT
correspondence rely on the supergravity approximation to the
full string theory.

Recently, a third maximally
symmetric 10-dimensional solution of
type IIB string theory was found \cite{Blau:2001ne,Blau:2002dy}:
the so-called maximally symmetric
plane wave background\footnote{This background is sometimes
called the `pp-wave'. This imprecise term has been adopted in most
of the literature on the subject.}.
This background can be obtained
from $AdS_5 \times S^5$ by taking a Penrose limit, and it is essentially
flat space plus the first order correction from flat space to
the full $AdS_5 \times S^5$.
In some sense this background sits half-way between flat space
and $AdS_5 \times S^5$.
It resembles $AdS_5 \times S^5$ because it is a curved geometry
with non-zero five-form flux, and it resembles flat space because
remarkably, the free string theory in this background
can be solved exactly in light-cone gauge \cite{Metsaev:2001bj}
(the details will be presented in Lecture~1).

Furthermore it was realized in \cite{Berenstein:2002jq} that
the Penrose limit of $AdS_5 \times S^5$ has a very simple description
in terms of the dual Yang-Mills theory.
In particular, BMN related
IIB string theory on the plane wave background to
a sector of the large $N$ limit
of ${\cal{N}}=4$ SU($N$) gauge theory involving
operators of large R-charge $J \sim \sqrt{N}$.

Because string theory in the plane wave is exactly solvable,
the BMN correspondence opens up the exciting opportunity
to study stringy effects in the holographic dual gauge theory,
thereby adding a new dimension to our understanding of gauge
theory, gravity, and the holographic principle.
Light-cone string field theory, although somewhat esoteric, has emerged
from a long hibernation
because it naturally connects the string and gauge theory descriptions of
the plane wave limit of AdS/CFT.

\subsection{What is and what is not in these Notes}

Light-cone string field theory is
a very complicated and technical subject.   In order to
provide a pedagogical introduction, most of
the material is developed for the simpler case of the
bosonic string, although we do mention the qualitatively
new features which arise for the type IIB string.
The reader who is interested in a detailed study of the latter,
significantly more complicated case is encouraged
to refer to the papers
\cite{Spradlin:2002ar,Spradlin:2002rv,Pankiewicz:2002tg},
or to review articles
such as \cite{Plefka:2003nb,Pankiewicz:2003pg}.
Our goal here has been to present in a clear
manner the relevant background
material which is usually absent in the recent literature
on plane waves.

\section{Lecture 1: The Plane Wave Limit of AdS/CFT}

\subsection{Lightning Review of the AdS/CFT Correspondence}

We start with a quick review of the remarkable duality between
${\cal{N}}=4$ SU($N$) super-Yang Mills theory and
type IIB string theory on $AdS_5 \times S^5$, referring
the reader to \cite{Aharony:1999ti} for an
extensive review and for additional references.
This is the best understood and most concrete example
of a holographic duality between gauge theory and string theory.

The Lagrangian for ${\cal{N}} = 4$ super-Yang Mills theory is
\cite{D'Hoker:2002aw}
\begin{equation}
\label{eq:lagrangian}
{\cal{L}} = 
{\rm Tr} \left[
- {1 \over 2 g_{\rm YM}^2}
F_{\mu \nu} F^{\mu \nu} - {1 \over 2}
\sum_i D_\mu 
\phi^i D^\mu \phi^i
+ {g_{\rm YM}^2 \over 4} \sum_{i, j} [ \phi^i, \phi^j]^2 
+ {\rm fermions}
\right],
\end{equation}
where $F_{\mu \nu}$ is the gauge field strength and $\phi^i$, $i=1,\ldots,6$
are six real scalar fields.  All fields transform in the adjoint
representation of the gauge group.
The ${\cal{N}} = 4$ superalgebra in four dimensions
is very constraining and
essentially determines the field content and the
Lagrangian~(\ref{eq:lagrangian})
uniquely:  the only freedom is the choice of gauge group
(which will always be SU($N$) in these lectures), and the value
of the coupling constant $g_{\rm YM}$.  This theory is conformally
invariant, so $g_{\rm YM}$ is a true parameter of the theory.
(In non-conformal theories, couplings become functions of
the energy scale, rather than parameters.)
The symmetry of the Lagrangian~(\ref{eq:lagrangian})
is SO(2,4)$\times$SO(6), where the first factor is the
conformal group in four dimensions and the second group
is the SO(6) R-symmetry group which acts on the six
scalar fields in the obvious way.

On the other side of the duality we have type IIB string
theory on the $AdS_5 \times S^5$ spacetime, whose metric can be
written as
\begin{equation}
\label{eq:metric}
ds^2 = {r^2 \over R^2} (-dx_0^2 + dx_1^2 + dx_2^2 + dx_3^2) + R^2 {dr^2 \over
r^2} + R^2 d\Omega_5^2.
\end{equation}
This is a solution of the IIB equations of motion with constant string
coupling $g_s$ and five-form
field strength
\begin{equation}
\label{eq:fiveform}
F_5 = (1 + *)
dx_0 \wedge dx_1 \wedge dx_2 \wedge dx_3 \wedge d(R^4/r^4).
\end{equation}
The isometry group of the spacetime is SO(2,4)$\times$SO(6), where
the first factor is the isometry of $AdS_5$ and the second factor
is the isometry of $S^5$.

According to the AdS/CFT correspondence, there is a one-to-one
correspondence between single-trace operators ${{O}}$ in the gauge theory
and fields $\phi$ in AdS.  The holographic dictionary between
the two sides of this duality 
is summarized in Table~\ref{TABLE1}, where $\lambda =
g_{\rm YM}^2 N$ is the 't Hooft coupling and $\sqrt{\alpha'}$ is
the string length.

\begin{table}[htb]
\begin{center}
\begin{tabular}{|ccc|} \hline
& & \\
${\cal{N}} = 4$ SU($N$) super-Yang Mills & $\Longleftrightarrow$
& IIB string theory on $AdS_5 \times S^5$ \\
& & \\
\hline
\hline
& & \\
$g_{\rm YM}^2$ & $=$ & $4 \pi g_s$ \\
& & \\
$\lambda^{1/4} = (g_{\rm YM}^2 N)^{1/4}$ & $=$ & $R/\sqrt{\alpha'}$ \\
& & \\
$\left< \exp \left[\int d^4 x\ \phi_0(x) {{O}}(x)\right] \right>_{\rm CFT}$
& $=$ & $Z_{\rm string}\left[\left.\phi(x,r)\right|_{r=0} = \phi_0(x) \right]$
\\
& & \\ \hline
\end{tabular}
\caption{The AdS/CFT correspondence for $AdS_5 \times S^5$. \label{TABLE1}}
\end{center}
\end{table}

\subsection{The Penrose Limit of $AdS_5 \times S^5$}

Now we consider a particular limit (a special case of a 
{\it Penrose Limit}) of the $AdS_5 \times S^5$ background,
following the treatment in \cite{Berenstein:2002jq}.
The limit we consider
can be thought of as focusing very closely upon the neighborhood
of a particle which is sitting in the `center' of $AdS_5$ and moving
very rapidly (close to the speed of light) along the equator of
the $S^5$.  To this end it is convenient to write the $AdS_5 \times S^5$
metric in  the
following coordinate system:
\begin{equation}
\label{eq:newmetric}
{ds^2 \over R^2} = \underbrace{-
dt^2 \cosh^2 \rho + d \rho^2 + \sinh^2 \rho d\Omega_3^2}_{AdS_5}
+ \underbrace{d\psi^2 \cos^2 \theta + d \theta^2 + \sin^2 \theta
d\Omega_3'^2}_{S^5}.
\end{equation}
Now $\rho = 0$ is the `center' of $AdS_5$ and $\rho = \infty$ is the
boundary.  (These are respectively $r=\infty$ and $r=0$ in
the coordinate system of~(\ref{eq:metric})).
The coordinate $\theta \in [- {\pi \over 2}, {\pi \over 2} ]$ is
the `latitude' on the $S^5$ and $\psi$, which is periodic modulo $2 \pi$,
is the coordinate along the equator of the $S^5$.

Note that by singling out an equator along the $S^5$, we have broken
the manifest SO(6) isometry of the metric on $S^5$
down to U(1)$\times$SO(4).

Now consider the coordinates $\widetilde{x}^\pm = {1 \over 2} (t \pm \psi)$,
which are appropriate for a particle traveling along the trajectory
$t - \psi \approx 0$.
To focus in on the neighborhood of this particle, which is sitting at
$\rho = \theta = 0$, means to
consider the following range of coordinates:
\begin{equation}
\widetilde{x}^+ = {\rm finite}, \qquad \widetilde{x}^- = {\rm infinitesimal}, \qquad
\rho = {\rm infinitesimal}, \qquad y = {\rm infinitesimal}.
\end{equation}
In order to isolate this range of coordinates,
it is convenient to rescale the coordinates according to
\begin{equation}
x^+ = \widetilde{x}^+, \qquad x^-= R^2 \widetilde{x}^-, \qquad
\rho = {r \over R}, \qquad \theta = {y \over R}
\end{equation}
and then take the limit $R \to \infty$.\footnote{Think of the $R 
\to \infty$ limit as the $N \to \infty$ limit in the dual
gauge theory; we will be keeping
$g_{\rm YM}$ fixed.}
Taking this
limit of~(\ref{eq:newmetric}) brings the metric into the form
\begin{equation}
ds^2 = - 4 dx^+ dx^- - (r^2 + y^2) (dx^+)^2 +dr^2+ r^2 d\Omega_3^2 +
dy^2 + y^2 d\Omega_3'^2.
\end{equation}
The last four terms are just the flat metric
on ${\mathbb{R}}^4 \times {\mathbb{R}}^4= {\mathbb{R}}^8$,
so we can rewrite the metric 
more simply as
\begin{equation}
\label{eq:ppmetric}
ds^2 = - 4 dx^+ dx^- -  \mu^2 x^2 (dx^+)^2 + dx_I dx^I,
\end{equation}
where $I=1,\ldots,8$.
In this formula we have introduced a new parameter $\mu$.
Note that $\mu$ is essentially irrelevant since it can always be
eliminated by a Lorentz boost in the $x^+$--$x^-$ plane, $x^\pm \to
x^\pm \mu^{\pm 1}$.  However, $\mu$ will serve as a useful bookkeeping
device.

We should not forget about the five-form field strength~(\ref{eq:fiveform}),
which remains
non-zero in the Penrose limit of the $AdS_5 \times S^5$ solution.
Taking the appropriate limit of~(\ref{eq:fiveform}) gives
\begin{equation}
\label{eq:ppform}
F_{+1234} = F_{+5678} = {\mu \over 4 \pi^3 g_s \alpha'^2}.
\end{equation}

The metric~(\ref{eq:ppmetric}) and five-form~(\ref{eq:ppform})
themselves
constitute the maximally symmetric plane wave
(`pp-wave') solution of the equations of motion
of type IIB string theory.  Note that the full symmetry of
this background is SO(4)$\times$SO(4)$\times {\mathbb{Z}}_2$.  The first SO(4)
is a remnant of the SO(2,4) isometry group of $AdS_5$ and the second
SO(4) is a remnant of the SO(6) isometry group of $S^5$.  The ${\mathbb{Z}}_2$
symmetry exchanges these two SO(4)'s, acting on the coordinates
$x^i$ by $(x_1,x_2,x_3,x_4) \leftrightarrow (x_5,x_6,x_7,x_8)$.
This peculiar discrete symmetry survives only in the strict pp-wave limit.
This symmetry is broken if we perturb slightly away from the limit back
to $AdS_5 \times S^5$.

In summary:  we started with the $AdS_5 \times S^5$ solution
of IIB string theory, and we took a limit which focused on 
the neighborhood around a trajectory traveling very rapidly around
the equator of the $S^5$, and we arrived at a different solution
of IIB string theory.
The natural question is now:  what does this Penrose limit correspond
to on the gauge theory side of the AdS/CFT correspondence?

First, note that since we had to break the SO(6) symmetry of the
$S^5$ by choosing an equator, then on the gauge theory side we
must also break the SO(6) symmetry to SO(4)$\times$U(1) by choosing
some U(1) subgroup of the R-symmetry group.  Without loss of
generality we can choose this U(1) subgroup to be the group
of rotations in the $\phi^5$--$\phi^6$ plane.  From
now on, when we talk about the R-charge of some state,
we mean the charge of the state with respect to this U(1) subgroup
of the full R-symmetry group.

Next, it is useful to trace through the above
coordinate transformations to see what the light-cone energy $p^-$ and
light-cone momentum $p^+$
correspond to on the gauge theory side.\footnote{Caution:  It has
become standard in the literature to {\bf define} $2 p^+ =
- p_-$, rather than to use the inverse metric (which has
$g^{--} \ne 0$) to raise the indices.}
To this end, recall that the energy in global coordinates in
$AdS_5$ is given by $E = i \partial_t$
and the angular momentum (around the equator of the $S^5$)
is $J = -i \partial_\psi$.
In terms of the dual CFT, these correspond respectively to the
conformal dimension $\Delta$ and R-charge of an operator.
Therefore 
we obtain the identification:
\begin{eqnarray}
\label{eq:eandm}
&& 2 p^- = - p_+ = i \partial_{x^+} = i \partial_{\widetilde{x}^+}
= i (\partial_t + \partial_\psi) = \Delta - J,
\\
&& 2 p^+ \equiv - p_- = - {\widetilde{p}_- \over R^2}
= {1 \over R^2} i \partial_{\widetilde{x}^-}
= {1 \over R^2} i (\partial_t - \partial_\psi) = {\Delta + J \over R^2}.
\label{eq:pplus}
\end{eqnarray}

On the string theory side, when we say that we `focus in on' a
small neighborhood of the equatorial trajectory, what that means
is that we consider amongst all the possible fluctuations of
IIB string theory on $AdS_5 \times S^5$ only those which are localized
in that small neighborhood. 
Now~(\ref{eq:eandm}) and~(\ref{eq:pplus})
suggest that on the gauge theory side,
this truncation corresponds to considering those operators which have
finite $\Delta - J$, and $\Delta + J \sim R^2 \sim \sqrt{N}$.
Therefore we arrive at the 
so-called BMN correspondence, as summarized in Table~\ref{TABLE2}.

\begin{table}[htb]
\begin{center}
\begin{tabular}{|ccc|} \hline
& & \\
${\cal{N}} = 4$ SU($N$) super-Yang Mills,  & $
\Longleftrightarrow$ & IIB string theory on
$AdS_5 \times S^5$,\\
in the limit $N \to \infty$, $g_{\rm YM} = $ fixed,
&& in the limit $R \to \infty$, $g_s = $ fixed,\\
truncated to operators with && truncated to states with \\
$\Delta \sim J \sim \sqrt{N}$ and finite $\Delta - J$
&& finite $p^+$ and $p^-$
\\ 
& & \\
$||$ & & $||$ \\ 
& & \\
? & & IIB string theory on a plane wave \\ 
& & \\ \hline
& & \\
$\Delta - J$ &$=$ & ${2\over \mu} p^-$ \\
& & \\
$J/\sqrt{\lambda}$ & $=$& $\mu p^+ \alpha'$
\\
& & \\ \hline
\end{tabular}
\caption{The plane wave limit of the AdS/CFT correspondence. \label{TABLE2}}
\end{center}
\end{table}

It will be very convenient to define the quantities
\begin{equation}
\lambda' = {g_{\rm YM}^2 N \over J^2} = {1 \over (\mu p^+ \alpha')^2}, 
\qquad  g_2 = {J^2 \over N}
\end{equation}
which remain finite in the BMN limit.  Also, we will refer to those
operators with $\Delta \sim J \sim \sqrt{N}$ and finite $\Delta  - J$
as `BMN operators'.\footnote{Some papers use the term `BMN operator'
strictly for those which are non-BPS.
We will use the term
`BMN operator' inclusively to include even BPS operators which survive
in the BMN limit.}

In Table~\ref{TABLE2} we have introduced the parameter
$\mu$ by rescaling $x^\pm$ as discussed above.
The question mark in Table~\ref{TABLE2} indicates that we are still
looking for a nice way to characterize this limit of
the gauge theory.  In other words, what precisely does
it mean to `truncate' the theory to a certain class of operators;
or, turned around:  is there a simple description of the sector
of the gauge theory which is dual to IIB string theory on a plane wave?

\subsection{Strings on Plane Waves}

The most exciting aspect of the BMN correspondence is that the free IIB string
on the plane wave background is exactly solvable.
As discussed in the introduction,
string theory on $AdS_5 \times S^5$ is in contrast rather complicated.

In light cone gauge, the worldsheet theory for IIB strings on
the plane wave background (Green-Schwarz action)
is simply \cite{Metsaev:2001bj}
\begin{equation}
\label{eq:wsact}
S = {1 \over 2 \pi \alpha'} \int dt
\int_0^{2 \pi \alpha' p^+} d \sigma \left[
{1 \over 2} \dot{X}^2 - {1 \over 2} X'^2 - {1 \over 2} \mu^2 X^2
+i \overline{S} (\partial\!\!\!\slash + \mu \Pi) S \right],
\end{equation}
where $X^I$, $I=1,\ldots,8$ are the bosonic sigma model coordinates,
$S$ is a complex Majorana spinor on the worldsheet and a positive
chirality SO(8) spinor under rotations in the transverse directions,
and $\Pi = \Gamma^1 \Gamma^2 \Gamma^3 \Gamma^4$.

The action~(\ref{eq:wsact}) simply describes
eight massive bosons and eight massive fermions, so it is trivially
solvable.
Let us consider here only bosonic excitations.
Then a general state has the form
\begin{equation}
\label{eq:state}
a_{n_1}^{I_1} \cdots a_{n_m}^{I_m} |0; p^+\rangle,
\end{equation}
and the Hamiltonian
be written as
\begin{equation}
\label{eq:spectrum}
2 p^- = - p_+ = \sum_{n=-\infty}^\infty  \sum_{I=1}^8 
(a_{n}^I)^\dagger a_{n}^I \sqrt{\mu^2 + {n^2
\over (\alpha' p^+)^2}}.
\end{equation}
We have chosen a basis of Fourier modes such that $n>0$ label left movers,
$n<0$ label right movers, and $n=0$ is the zero mode.
This convention has become standard in the pp-wave literature and
contrasts with the usual convention in flat space, where
the left- and right-moving oscillators are denoted
by different symbols $\alpha_n$ and $\tilde{\alpha}_n$.

The alternate convention can be motivated by recalling that in
flat space, the worldsheet theory remains a conformal field theory
even in light-cone gauge.  Therefore the left-moving modes $\alpha_n$
and the right-moving modes $\tilde{\alpha}_n$ decouple from each other.
However, in the plane-wave background, choosing light-cone gauge breaks
conformal invariance on the world sheet (because a mass term appears).
Therefore all of the modes couple to each other, so there is no advantage
to introducing a notation which treats left- and right-movers separately.

Since this is a theory of closed strings, we should not forget
to impose the physical state condition, which says that the total
momentum on the string should vanish:
\begin{equation}
P = \sum_{n=-\infty}^\infty\sum_{I = 1}^8 n N^I_n = 0,
\end{equation}
where $N^I_n$ is the occupation number (the eigenvalue
of $(a_n^I)^\dagger a_n^I$).

We remarked above that the SO(8) transverse symmetry
is broken by the five-form field strength
to SO(4)$\times$SO(4)$\times {\mathbb{Z}}_2$.  This manifests itself by the
presence of $\Pi$ in the worldsheet action~(\ref{eq:wsact}).
However, the free spectrum of IIB string theory on the plane wave is
actually fully SO(8) symmetric \cite{Dobashi:2002ar}.
One can see this by noting
that $\Pi$ can be eliminated from~(\ref{eq:wsact}) by first
splitting $S = S_1 + i S_2$ and then making the field redefinition
$S_2 \to \Pi S_2$.  One should think of SO(8) as an accidental
symmetry of the free theory.  Since the background breaks
SO(8) to SO(4)$\times$SO(4)$\times {\mathbb{Z}}_2$, there is no reason
to expect that string interactions should preserve the full
SO(8), and indeed we will see that they do not:  the interactions
break SO(8) to SO(4)$\times$SO(4)$\times {\mathbb{Z}}_2$.

Before we return to the gauge theory, it will be convenient to
rewrite the spectrum~(\ref{eq:spectrum}) in terms of gauge theory
parameters using the dictionary from Table~\ref{TABLE2}:
\begin{equation}
\label{eq:gspec}
\Delta - J = \sum_n  \sum_{I=1}^8 N^I_n \sqrt{1 + \lambda' n^2}.
\end{equation}

\subsection{Strings from ${\cal{N}}=4$ Super Yang Mills}

We have seen that IIB string theory on the plane wave background
has a very simple spectrum~(\ref{eq:gspec}).  In this section we 
will recover this
spectrum by finding the set of operators which have
$\Delta \sim J \sim \sqrt{N}$ and finite $\Delta - J$ in the
$N \to \infty$ limit \cite{Berenstein:2002jq}.

Consider first the ground state of the string, $|0;p^+\rangle$.
According to~(\ref{eq:gspec}) 
this should correspond to an operator with $\Delta - J = 0$.
The unique such operator is ${\rm Tr}[Z^J]$, where
\begin{equation}
Z = {1 \over \sqrt{2}} (\phi_5 + i \phi_6)
\end{equation}
(recall that we defined $J$ to be the U(1) generator which
acts by rotation in the $\phi_5$--$\phi_6$ plane).
The first entry in the `BMN state-operator correspondence' is
therefore
\begin{equation}
|0;p^+\rangle \qquad \Longleftrightarrow \qquad 
 {\rm Tr}[Z^J].
\end{equation}

To get the first excited states we can add to the trace operators
which have $\Delta - J = 1$; for example:  $\phi_i$ (for $i=1,2,3,4$)
or $D_i Z$ (again for $i=1,2,3,4$, in the Euclidean theory).
\begin{eqnarray}
a_0^{i} |0; p^+ \rangle
& \Longleftrightarrow & {\rm Tr}[ \phi_i Z^J],
\cr
a_0^{i+4}
|0; p^+ \rangle & \Longleftrightarrow & {\rm Tr}[
D_i Z  Z^J].
\end{eqnarray}

To get higher excited states we can add more `impurities' to the
traces.  For example, a general state with $\Delta - J = k$ is
\begin{equation}
a_0^{i_1} a_0^{i_2}
\cdots 
a_0^{i_k}  |0;p^+\rangle  \Longleftrightarrow 
\sum {\rm Tr} [ \cdots Z \phi_{i_1}  Z \cdots 
Z \phi_{i_k}  Z \cdots ].
\end{equation}
The sum on the right hand side runs over all possible orderings
of the insertions inside the trace.  This sum is necessary to ensure
that the operator is BPS.
We will always work in a `dilute gas' approximation, where the number
of impurities is much smaller than $J$, the number of $Z$'s.

So far we considered only BPS operators in the gauge theory.
These have the property that $\Delta$ is not corrected
by interactions, i.e.~$\Delta$ does not depend on $g_{\rm YM}$.
According to~(\ref{eq:gspec}), these can only correspond to
string states with the zero mode ($n=0$) excited.
In order to obtain other states, we can consider  summing
over the location of an impurity with a phase:
\begin{equation}
a_n^i |0;p^+\rangle \qquad \Longleftrightarrow \qquad
 \sum_{k=0}^J e^{2 \pi i n k/J} {\rm Tr}[ Z^k \phi_i Z^{J-k} ].
\end{equation}
But the right hand side is zero (for $n \ne 0$),
because of cyclicity of the trace!
Actually  this is a good thing, because the string state on
the left does not satisfy the physical state condition for $n \ne 0$.

In order to get physical states, we have to consider 
(suppressing the $i$ transverse index)
\begin{equation}
\label{eq:genstate}
a_{n_1} \cdots a_{n_m}
|0;p^+\rangle  \Longleftrightarrow
 \sum_{k_1,\ldots,k_m=0}^J e^{2 \pi i (n_1 k_1 + \cdots +n_m k_m)/J}
{\rm Tr}[ \cdots Z \phi Z \cdots Z \phi Z\cdots ].
\end{equation}
Here $k_i$ labels the position, in the string of $Z$'s, of the
$i$-th $\phi$ impurity.
Cyclicity of the trace now implies that the right-hand side vanishes
unless $n_1 + \cdots +n_m = 0$, and this is precisely the physical
state condition for the string state on the left-hand side!

The operator on the right-hand side of~(\ref{eq:genstate}) is
not BPS when the phases are non-zero, so its dimension $\Delta$
receives quantum corrections.  One can check
that in the BMN limit $N,J \to \infty$ with $J \sim \sqrt{N}$,
the contribution to $\Delta - J$ from an impurity with
phase $n$ is
\begin{equation}
(\Delta - J)_n = \sqrt{1 + \lambda' n^2},
\end{equation}
precisely in accord with the prediction~(\ref{eq:gspec})!
This calculation was performed to one loop in \cite{Berenstein:2002jq},
to two loops in \cite{Gross:2002mh}, and an argument
valid to all orders in perturbation theory was presented
in \cite{Santambrogio:2002sb}.

At this point we have motivated that the spectrum of IIB string theory
on the plane wave background can be identified with the set
of BMN operators in ${\cal{N}}=4$ SU($N$) Yang-Mills theory.
One point which we did not consider is the addition of impurities
with $\Delta - J > 1$, for example $\overline{Z}$, which
has $\Delta -J = 2$.  It has been argued that these operators
decouple in the BMN limit (i.e., their anomalous dimensions go to
infinity), and can hence be ignored.  We refer the reader
to \cite{Berenstein:2002jq} for details.
Next we present the two parameters \cite{Kristjansen:2002bb,
Constable:2002hw} which characterize the BMN limit.

\subsection{The Effective 't Hooft Coupling $\lambda'$}

Something a little miraculous has happened.  The operator~(\ref{eq:genstate})
is not BPS (when the phases are nonzero), so its conformal
dimension $\Delta$ will receive quantum corrections.  Generically,
the dimension of a non-protected operator blows up 
in limit of large 't Hooft coupling.
And we certainly are taking $\lambda  = g_{\rm YM}^2 N
\to \infty$ here (see
Table~\ref{TABLE2})!

However, note that~(\ref{eq:genstate})
is BPS when all of the phases are zero:  $n_1 = \cdots = n_m = 0$.
In a sense, then, we might hope that the operator is `almost' BPS
as long as the phases are almost 0,
or in other words $n_i/J \ll 1$ for all $i$.  
By `almost' BPS we mean that although the dimension does 
receive quantum corrections, those corrections are finite
in the BMN limit despite the fact that $\lambda \to \infty$.
Indeed this is what happens:  in the formula~(\ref{eq:gspec}) 
is it not the `t Hooft coupling $\lambda$ which appears, but
rather a new effective coupling $\lambda' = \lambda/J^2$ which is
finite in the BMN limit.

It is hoped (and indeed this hope is borne out in all known
calculations so far) that this miracle is quite general in
the BMN limit:  namely, that many interesting
physical quantities related to these
BMN operators remain finite despite the fact that the `t Hooft
coupling is going to infinity.

\subsection{The Effective Genus Counting Parameter $g_2$}

In the familiar large $N$ limit of SU($N$) gauge theory, the
perturbation theory naturally organizes into a genus expansion,
where a gauge theory diagram of genus $g$ is effectively weighted
by a factor of $1/N^{2 g}$.  In particular, only planar ($g=0$)
diagrams contribute to leading order at large $N$.

However, it has been shown \cite{Kristjansen:2002bb,
Constable:2002hw} that the effective genus counting parameter
for BMN operators
is $g_2 = J^2/N$.
The familiar intuition that only planar graphs are
relevant at $N=\infty$ fails because we are not focusing our attention
on some fixed gauge theory operators, and then taking $N \to \infty$.
Rather, the BMN operators themselves change with $N$, since we want
to scale the R-charge $J \sim \sqrt{N}$.  As $N$ becomes large,
the relevant BMN operators are composed of ${\cal{O}}(\sqrt{N})$
elementary fields.  Because of this large number of fields,
there is a huge number ${\cal{O}}(J^{4 g})$ of Feynman diagrams at genus $g$.
This combines with the $1/N^{2 g}$ factor to give a finite
weight $g_2^{2 g}$ in the BMN limit for genus $g$ diagrams.

\section{Lecture 2: The Hamiltonian of String Theory}

In this lecture we will canonically
second-quantize string theory in light-cone
gauge and write down its Hamiltonian, which will be no more complicated
(qualitatively) than
\begin{equation}
H = a^\dagger a + g_s (a^\dagger a a + a^\dagger a^\dagger a),
\end{equation}
where $g_s$ is the string coupling.
Students of string theory these days are not typically taught
that it is
possible to write down an explicit formula
for the Hamiltonian of string theory.
An excellent collection of papers on this subject may be found
in \cite{Schwarz}.
The light-cone approach does suffer from a number of problems which will be
discussed in detail in the next lecture.

However, light-cone string field theory is very well-suited to the study
of string interactions in the plane wave background.
For one thing, it is only in the light-cone gauge that we
are able to determine the spectrum of the free string.
Since other approaches cannot yet even give us the free spectrum,
they can hardly tell us anything about string interactions.
Although we hope this situation will improve, light-cone 
gauge is still very natural from the point of view of the BMN correspondence.
The dual BMN gauge theory automatically provides us a light-cone
quantized version of the string theory, and it is hoped that
taking the continuum limit of
the `discretized strings' in the gauge theory might give us light-cone string field
theory, although a large number of obstacles need to be overcome
before the precise correspondence is better understood.

Many of the fundamental concepts which will be introduced in this
and the following lecture apply equally well to all string field
theories, and not just the light-cone version.  It is therefore hoped
that these lectures may be of benefit even to some
who are not particularly interested in plane waves.

This portion of the lecture series is intended to be highly pedagogical.
We will therefore start
by studying the simplest possible case in great detail.
Here is a partial list of simplifications which we will start with:

\begin{itemize}
\item During this lecture we will consider only bosons.
Somewhat surprisingly,
fermions
complicate the story considerably,
but we will postpone these important details until the next lecture.
The reader should keep in mind that the plane wave background is not
a solution of bosonic string theory, so strictly speaking all of the formulas
presented in this lecture need to be supplemented by the appropriate 
fermions.
(The flat space limit $\mu \to 0$ does make sense without
fermions, if one works in 26 spacetime dimensions rather than 10.)

\item Since the bosonic sector of string theory on the plane wave background
is SO(8) invariant, we will completely ignore the transverse index
$I=1,\ldots,8$ for most of the discussion.  It is trivial 
to replace, for example, $x^2 \to \sum_{I=1}^8 (x^I)^2$ in all of the formulas
below.

\item Finally, we will begin not even with a string in the plane wave background,
but simply a particle in the plane wave background!  A string can essentially
be thought of as an infinite number of particles, one for each Fourier mode
on the string worldsheet.  A particle is equivalent
to taking just the zero-mode on the worldsheet (the mode independent of $\sigma$).
In flat space, there is a qualitative difference between this zero-mode and
the `stringy' modes:  the former has a continuous spectrum (just the overall
center-of-mass
center-of-mass
momentum of the string) while the stringy modes are like harmonic oscillators
and have a discrete spectrum.  However in the plane wave background, even
the zero-mode lives in a harmonic oscillator potential, so it is not qualitatively
different from the non-zero modes.  Once we develop all of the formalism
appropriate for a particle, it will be straightforward to take infinitely
many copies of all of the formulas and apply them to a string.

\end{itemize}

\subsection{Free Bosonic Particle in the Plane Wave Background}

We consider a particle propagating in the plane wave metric
\begin{equation}
ds^2 = 
- 2 dx^+ dx^- -\mu^2 x^2 (dx^+)^2 + dx^2.
\end{equation}
The action for a free `massless' field is
\begin{equation}
S = - {1 \over 2} \int \partial_\mu \Phi \partial^\mu \Phi =
\int dx^+ dx^- dx\ \partial_+ \Phi \partial_- \Phi - \int dx^+ H,
\end{equation}
where we have defined
\begin{equation}
H = {1 \over 2} \int dx^- dx \left[ (\partial_x \Phi)^2 + \mu^2
x^2 (\partial_- \Phi)^2 \right].
\end{equation}
Now let us canonically quantize this theory, so we promote $\Phi$
from a classical field to an operator on the multi-particle Hilbert
space.
The canonically conjugate field to $\Phi$ is $\partial_- \Phi$, so the
commutation relation is
\begin{equation}
[\Phi(x^-, x), \partial_- \Phi(y^-, y)] = i \delta(x^- - y^-) \delta(x-y).
\end{equation}
Let us pass to a Fourier basis by introducing
\begin{equation}
\Phi(x^-, x) = {1 \over 2 \pi} \int dp_- dp \ \Phi(p_-, p)
e^{i ( p_- x^- + p x)}.
\end{equation}
Note:  from now on we {\bf define}
\begin{equation}
p^+ = - p_-.
\end{equation}
In the supersymmetric string theory to be considered below, the supersymmetry
algebra will guarantee that $p^+ \ge 0$ for all states.
We will proceed with this assumption, although it is certainly not
true in the 26-dimensional bosonic theory.
The commutation relation is now
\begin{equation}
\label{eq:phiccr}
[\Phi(p^+, p), \Phi(q^+, q) ] = {1 \over p^+}
\delta(p^+ + q^+) \delta(p+q).
\end{equation}
Since $\Phi$ is real (as a classical scalar field), the corresponding
operator $\Phi$ is Hermitian, which means that
\begin{equation}
\label{eq:reality}
\Phi(p^+, p)^\dagger = \Phi(-p^+, -p).
\end{equation}
The Hamiltonian can now be written as
\begin{equation}
\label{eq:hamil}
H_2 = {1 \over 2} \int dp_- dp\ \Phi^\dagger ( p^2 +(\mu p^+ x)^2 ) \Phi =
\int dp^+ dp \ p^+ \Phi^\dagger h \Phi,
\end{equation}
(normal ordering will always be understood)
where $h$ is the single-particle Hamiltonian
\begin{equation}
h = {1 \over 2 p^+} (p^2 + \omega^2 x^2), \qquad \omega = \mu| p^+|.
\end{equation}
(We will always take $\mu \ge 0$.)
The subscript ``2'' in~(\ref{eq:hamil}) denotes that this is the quadratic
(i.e., free) part of the Hamiltonian.  Later we may add higher order interaction
terms.
The single-particle Hamiltonian may be diagonalized in the standard way:
\begin{equation}
a = {1 \over \sqrt{2 \omega}} (p - i \omega x), \qquad
p = \sqrt{\omega \over 2} (a + a^\dagger), \qquad x = {i \over \sqrt{2 \omega}}
(a - a^\dagger),
\end{equation}
so that
\begin{equation}
h = e(p^+) \mu ( a^\dagger a + {\textstyle{1 \over 2}}),
\end{equation}
where $e(x) = {\rm sign}(x)$.

It is important to distinguish two different Hilbert spaces.
The single-particle Hilbert space ${\cal{F}}$ is spanned by the vectors
\begin{equation}
|N; p^+ \rangle \equiv (a^\dagger)^N |0; p^+\rangle, \qquad N=0,1,\ldots.
\end{equation}
The operators $a$, $a^\dagger$ and $h$ act on ${\cal{F}}$.
The second Hilbert space is the multi-particle Hilbert space ${\cal{H}}$.
Let us introduce particle creation/creation operators $A_N(p^+)$ which
act on ${\cal{H}}$ and satisfy $(A_N(p^+))^\dagger = A_N(-p^+)$ and
\begin{equation}
\label{eq:accr}
[ A_M(p^+), A_N(q^+)] = e(p^+) \delta_{MN} \delta(p^+ + q^+).
\end{equation}
For $p^+>0$, $A_N(p^+)$ annihilates a particle in the state
$|N; p^+\rangle$, while for $p^+<0$, $A_N(p^+)$ creates a particle
in the state $|N; -p^+\rangle$.
The vacuum of ${\cal{H}}$,
denoted by $|0\rangle\!\rangle$, is annihilated by all $A_N(p^+)$ which have
$p^+>0$.
Normally in scalar field theory we do not introduce
this level of complexity because the `internal' Hamiltonian $h$
is so trivial.

We may refer to ${\cal{F}}$ as the `worldsheet' Hilbert space
and ${\cal{H}}$ as the `spacetime' Hilbert space, since this is of course
how these should be thought of in the string theory.

We now write the usual expansion for $\Phi$ in terms of particle creation
and annihilation operators:
\begin{equation}
\label{eq:phifield}
\Phi(p^+) = {1 \over \sqrt{|p^+|}} \sum_{N=0}^\infty |N;p^+\rangle A_N(p^+).
\end{equation}
It is easily checked that the commutation relation~(\ref{eq:phiccr})
follows from~(\ref{eq:accr}).
Note that we have written $\Phi$ as simultaneously a state in ${\cal{F}}$
and an operator in ${\cal{H}}$.
This is notationally more convenient than the position space representation,
\begin{equation}
\Phi(p^+,x) =
\langle x | \Phi(p^+) \sim
{1 \over \sqrt{|p^+|}} \sum_{N=0}^\infty e^{-x^2} H_N(x) A_N(p^+),
\end{equation}
which would leave all of our formulas full of Hermite polynomials $H_N(x)$.

Writing the field operator $\Phi$ as a state in the single-particle Hilbert
space has notational advantages other than just being able to do without
Hermite polynomials.  For example, the equation of motion
\begin{equation}
\partial_+ \partial_- \Phi - {1 \over 2} \partial^2_x \Phi - {1 \over 2}
\mu^2 x^2 \partial_-^2 \Phi = 0
\end{equation}
is just the Schr\"odinger equation
on $\Phi$:
\begin{equation}
i \partial_+ \Phi = h \Phi.
\end{equation}

There is
also a simple formula which allows us to take any symmetry generator
on the worldsheet (such as $h$ above, and later rotation generators $j^{IJ}$,
supercharges $q$, $\ldots$) and construct a free field realization of the
corresponding space-time operator ($H$, $J^{IJ}$, $Q$, $\ldots$):
\begin{equation}
\label{eq:Gdef}
G_2 = \int dp^+ dp \ p^+ \Phi^\dagger g \Phi.
\end{equation}
We already saw this formula applied to the Hamiltonian in~(\ref{eq:hamil}).
When we further make use of the expansion of $\Phi$  into creation operators,
we find the expected formula
\begin{equation}
H_2 = \int_0^\infty dp^+ \sum_{N=0}^\infty E_N A_N(-p^+) A_N(p^+),\qquad
E_N = \mu (N + {\textstyle{1 \over 2}}).
\end{equation}
Another trivial application is the identity operator --- it is easily checked
that
\begin{equation}
\int dp^+\ p^+ \Phi^\dagger \Phi=
- i \int dx^- \ \Phi \partial_- \Phi =
\int_0^\infty dp^+ \sum_{N=0}^\infty A_N(-p^+) A_N(p^+) = I.
\end{equation}
Again, the subscript `2' in~(\ref{eq:Gdef}) emphasizes that this gives
a free field realization (i.e., quadratic in $\Phi$).  Dynamical symmetry
generators (such as the Hamiltonian, in particular) will pick up additional
interaction terms, but kinematical symmetry generators (such as $J^{IJ}$)
remain quadratic.

\subsection{Interactions}

Now let us consider a cubic interaction, 
\begin{equation}
H_3 = g_s \int dx^- dx \ V,
\end{equation}
where, for example, we might choose
\begin{equation}
V = \Phi^3 +
\Phi (\partial_-^{17} \Phi) (\partial_x^6 \Phi).
\end{equation}
If we insert the expansion of $\Phi$ in terms of modes and do
the integrals,
we end up with an expression of the form
\begin{equation}
\label{eq:eqaa}
H_3 = \int dp_1^+ dp_2^+ dp_3^+ \delta(p_1^+ + p_2^+ +
p_3^+)\sum_{N,P,Q=0}^\infty
c_{NPQ}(p_1^+, p_2^+, p_3^+)
A_N(p_1^+)
A_P(p_2^+)
A_Q(p_3^+).
\end{equation}
The functions $c_{NPQ}(p_1^+, p_2^+, p_3^+)$ are obtained
from $V$ without too much difficulty:
they simply encode the matrix
elements of the interaction written in a basis of harmonic
oscillator wavefunctions.

We now adopt a convention which is important to keep in mind.
Because of the $p^+$ conserving delta function, it will always be the
case that two of the $p^+$'s are positive and one is negative, or vice versa.
We will always choose the index `3' to label the $p^+$ whose sign is opposite
that of the other two.
This means that the particle labeled
`3' will always be the initial state of a splitting transition
$3 \to 1 + 2$ or the final state of a joining transition
$1 + 2 \to 3$.

In the state-operator correspondence, it is convenient to identify the
operator cubic $H_3$ with a state in the 3-particle Hilbert space $|V\rangle$
(where $V$ stands for `vertex')
with the property that
\begin{equation}
\qquad \langle N; p_1^+| \langle P; p_2^+| \langle Q; -p_3^+|
H_3 \rangle = c_{NPQ}(p_1^+,p_2^+,p_3^+), \qquad {\rm for~} p_1^+, p_2^+ > 0, p_3^+<0.
\end{equation}
This state can be constructed by taking
\begin{equation}
\label{eq:eqbb}
|V\rangle = \sum_{N,P,Q=0}^\infty
c_{NPQ}(p_1^+,
p_2^+, p_3^+)|N;p_1^+ \rangle |P; p_2^+\rangle |Q;-p_3^+ \rangle.
\end{equation}

\bigskip
\noindent
{\bf Exercise.} Compute $c_{NPQ}(p_1^+,p_2^+,p_3^+)$ and $|V\rangle$
for $V(\Phi) = \Phi^3$.

\subsection{Free Bosonic String in the Plane Wave}

It is essentially trivial to promote all of the formulas from the preceding
section to the case of a string.
The field $\Phi(x)$ is promoted from a function of $x$, the position of the
particle in space, to a functional $\Phi[x(\sigma)]$ of the embedding
$x(\sigma)$ of the
string worldsheet in spacetime.  
In all of the above formulas, integrals over $dx$ are replaced by
functional integrals $ Dx(\sigma)$, and delta functions in $x$
are replaced by delta-functionals $\Delta[x(\sigma)]$.
These are defined as a product of delta functions over all of the Fourier
modes $x_n$ of $x(
\sigma)$.

The interacting quantum field theory of strings is described by the action
\begin{equation}
S = \int dx^+ dx^- Dx(\sigma) \ \partial_+ \Phi \partial_- \Phi 
- \int dx^+\ H,
\end{equation}
where $H = H_2 + H_3 + \cdots$.
The formula~(\ref{eq:Gdef}) is replaced by
\begin{equation}
\label{eq:gtwodef}
G_2 = \int dp^+  D p(\sigma)\ p^+  \Phi^\dagger g \Phi.
\end{equation}
The worldsheet Hamiltonian is now
\begin{equation}
h = {e(p^+) \over 2} \int_0^{2 \pi |p^+|}
d\sigma \left[ 4 \pi p^2
 + {1 \over 4 \pi} ( (\partial_\sigma x)^2 + \mu^2 x^2) \right]
= {1 \over p^+} \sum_{n=-\infty}^\infty \omega_n a_n^\dagger a_n,
\end{equation}
where
\begin{equation}
\omega_n = \sqrt{n^2 + (\mu \alpha' p^+)^2}
\end{equation}
is the energy of the $n$-th mode and we have
introduced a suitable
basis of raising and lowering operators in order to diagonalize $h$.
Note that $a_0$ is identified with the operator $a$ corresponding to 
a particle,
while for $n>0$, 
$a_n = 
\alpha_n$ are the left-movers and $a_{-n} = \widetilde{\alpha}_n$ are
the right-movers.

The full Hilbert space ${\cal{F}}$ of a single string is obtained
by acting on $
|0; p^+\rangle$ with the raising operators $a_n^\dagger$ (for all $n$!  Note
that we do not use any convention like $a_n^\dagger = a_{-n}$).
We therefore label a state by $|\vec{N}\rangle$, where the component
$N_n$ of the vector $\vec{N}$ gives the occupation number of oscillator $n$.
Note that we have to impose the
$L_0 - \overline{L}_0 = 0$ physical state condition
\begin{equation}
\sum_{n=-\infty}^\infty n N_n = 0.
\end{equation}

The second quantized Hilbert space ${\cal{H}}$ is introduced as before.
It has the vacuum $|0\rangle\!\rangle$, which is acted on by the
operators $A_{\vec{N}}(p^+)$, which for $p^+<0$
create a string in the state $|\vec{N}; p^+\rangle$.
The representation of the Hamiltonian at the level of free fields
is just
\begin{equation}
H_2 = \int_0^\infty dp^+ \sum_{|\vec{N} \rangle 
\in {\cal{F}}} E_{\vec{N}} A^\dagger_{\vec{N}}(p^+)
 A_{\vec{N}}^{} (p^+), \qquad
E_{\vec{N}} = {1 \over p^+} \sum_{n=-\infty}^\infty 
\omega_n N_n
\end{equation}

\subsection{The Cubic String Vertex}

Our goal now is to construct the state $|V\rangle$ in ${\cal{F}}^3$
which encodes the cubic string interactions, in the sense of formulas~(\ref{eq:eqaa})
and~(\ref{eq:eqbb}).  What is the principle that determines the cubic interaction?
It is quite simple:  the embedding of the string worldsheet into spacetime
should be continuous. 

In a functional representation, the cubic interaction is therefore just
\begin{equation}
\label{eq:split}
H_3 = g_2 \int 
 \delta(p_1^+ + p_2^+ + p_3^+)
f(p_1^+, p_2^+, p_3^+)
\Delta[ x_1(\sigma) + x_2(\sigma) - x_3(\sigma)]\prod_{r=1}^3
\left( dp_r^+ D x_r(\sigma)\ \Phi[ p_r^+, x_r(\sigma)]
\right).
\end{equation}
There is one very important caveat:  the principle of continuity
requires the delta functional $\Delta[ x_1(\sigma) + x_2(\sigma)
- x_3(\sigma)]$,
but it does not determine the interaction~(\ref{eq:split}) uniquely
because we have the freedom to choose the measure factor
$f(p_1^+, p_2^+, p_3^+)$ arbitrarily.  Moreover,
in principle the cubic interaction could involve derivatives of $\Phi$,
such as $\delta \Phi/\delta x(\sigma_{\rm I})$ (where $x_I$
is the interaction point), whereas the interaction we wrote
has only $\Phi^3$ with no derivatives.  We will return to these
points later.

\begin{figure}[htb]
\centerline{\hbox{ \psfig{figure=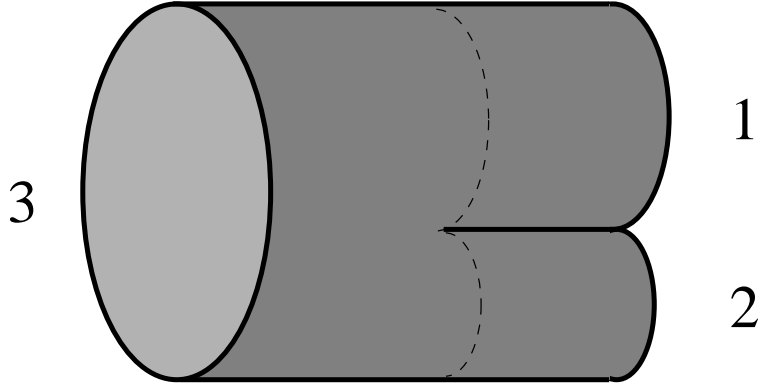} }}
\caption{\footnotesize The three-string interaction
vertex.  In order to make the picture clear we have drawn a finite
time interval enclosing the interaction, which occurs at the single
precise moment of time. Our goal
in this lecture is to write down the Feynman rule $|V\rangle$
for this simple interaction.  \label{fig:FIG1}}
\end{figure}

Our convention about the selection of $p_3^+$ guarantees that
string 3 is always the `long string'.
The interaction~(\ref{eq:split})
mediates the string splitting $3 \to 1 + 2$, or its hermitian conjugate,
the joining of $1 + 2 \to 3$.  This process is depicted
in Figure~\ref{fig:FIG1}.
All we have to do now is Fourier transform this delta-functional into
the harmonic oscillator number basis!  Let us assemble the steps of
this calculation.

\bigskip
\noindent
{\bf Step 1.}
First we recall the definition of a $\Delta$-functional as a product
of delta-functions for each Fourier mode,
\begin{equation}
\Delta[ x_1(
\sigma) + x_2(\sigma)  - x_3(\sigma)]
= \prod_{m=-\infty}^\infty \delta \left( \int_0^{2 \pi |p_3^+|}
d\sigma \ e^{i m \sigma/|p_3^+|} [
x_1(
\sigma) + x_2(\sigma)  - x_3(\sigma)] \right).
\end{equation}
Let us introduce matrices $X^{(r)}_{mn}$ which express the Fourier
basis of string $r$ in terms of the Fourier basis of string 3
(so that, clearly, $X^{(3)} = 1$).  Then we can write
\begin{equation}
\Delta[ x_1(\sigma) + x_2(\sigma) - x_3(\sigma)]
= \prod_{m=-\infty}^\infty \delta\left(
\sum_{r=1}^3 X^{(r)}_{mn} p_{n(r)} \right).
\end{equation}
These matrices are obtained by simple Fourier transforms,
\begin{equation}
X^{(1)}_{mn} = {1 \over \pi} (-1)^{m+n+1} {\sin(\pi m y) \over
n - m y}, \qquad
X^{(2)}_{mn} = {1 \over \pi} (-1)^n {\sin \pi m (1-y)  \over n - m (1-y)},
\end{equation}
where $y = p_1^+/|p_3^+|$ is the ratio of the width of string 1 to the
width of string 3.

\bigskip
\noindent
{\bf Step 2.}
The expansion of the field $\Phi$ in position space is given by
\begin{equation}
\label{eq:sphi}
\Phi[p^+, x(
\sigma)] = {1 \over \sqrt{|p^+|}}\sum_{\vec{N}}
A_{\vec{N}}(p^+) \prod_{n=-\infty}^\infty \psi_{N_n(x_n)},
\end{equation}
where $\psi_N(x) = \langle x|N\rangle$ is a harmonic oscillator
wavefunction for the $N$-th excited level.
When we plug~(\ref{eq:sphi}) into the cubic  action~(\ref{eq:split}),
we find that the coupling between the three strings labeled by
$\vec{N}_{(1)}$, $\vec{N}_{(2)}$ and $\vec{N}_{(3)}$ is simply
\begin{equation}
\label{eq:start}
c(\vec{N}_{(1)}, \vec{N}_{(2)}, \vec{N}_{(3)})
= \int \prod_{n=-\infty}^\infty \psi_{N_{(1) n}}(x_{n(1)})
\psi_{N_{(2) n}}(x_{n(2)})
\psi_{N_{(3) n}} (x_{n(3)})
d M
\end{equation}
where the measure is
\begin{equation}
dM = f D x_1(\sigma) D x_2(\sigma) D x_3(\sigma) \Delta[x_1(\sigma)+x_2(\sigma)
-x_3(\sigma)].
\end{equation}

\bigskip
\noindent
{\bf Step 3.}
The next step is to note  that
an $x$-eigenstate of an oscillator with frequency $\omega$ may
be represented as
\begin{equation}
|x \rangle= ({\rm constant}) \exp \left[- {\omega x^2 \over 4}
+ i \sqrt{2 \omega} a^\dagger -{1 \over 2}
a^\dagger a^\dagger \right]|0\rangle.
\end{equation}
It follows from this that
\begin{equation}
\label{eq:neat}
\sum_{N=0}^\infty  |N\rangle \psi_N(x) =
({\rm constant}) \exp \left[- {\omega x^2 \over 4}
+ i \sqrt{2 \omega} a^\dagger -{1 \over 2}
a^\dagger a^\dagger \right]|0\rangle.
\end{equation}
Note that the overall constant is irrelevant since we can
absorb it into $f$.

\bigskip
\noindent
{\bf Step 4.}
Now let us assemble the couplings $c$ into the state $|V\rangle$:
\begin{equation}
|V\rangle = \sum_{ \vec{N}_1, \vec{N}_2, \vec{N}_3}
= c(\vec{N}_{(1)}, \vec{N}_{(2)}, \vec{N}_{(3)} )
| \vec{N}_{(1)}\rangle
| \vec{N}_{(2)}\rangle
| \vec{N}_{(3)}\rangle.
\end{equation}
Using~(\ref{eq:start}) and~(\ref{eq:neat}), we arrive finally at
\begin{equation}
\label{eq:final}
|V\rangle
 = \int dM \exp \left[
\sum_{k=-
\infty}^\infty
\sum_{r=1}^3 \left( - { \omega_{k(r)}^2 \over 4} x_{k(r)}^2
- {1 \over 2} ( a_{k(r)}^\dagger)^2 + i \sqrt{2 \omega_{k(r)}}
x_{k(r)} a_{k(r)}^\dagger \right)
\right]
|0\rangle.
\end{equation}
The functional measure is just a product over all Fourier modes:
\begin{equation}
dM =  \prod_{m=-\infty}^\infty \left(
\sum_{r=1}^3 \sum_{n=-\infty}^\infty X^{(r)}_{mn} x_{n(r)} \right)
\prod_{r=1}^3 \prod_{k=-
\infty}^\infty dx_{k(r)}.
\end{equation}
The delta-functions allow us to replace all of the modes
of string 3 in terms of the modes of strings 1 and 2.
Then~(\ref{eq:final}) is just a Gaussian integral in the infinitely
many variables $x_{k(1)}$, $x_{k(2)}$, $k=-\infty,
\ldots,+\infty$!

\bigskip
\noindent
{\bf Step 5.}
The Gaussian integral is easily done, and we find
\begin{equation}
\label{eq:vertexone}
|V \rangle =  f(p_1^+, p_2^+, p_3^+)
(\det \Gamma)^{-1/2} 
\exp
\left[ {1 \over 2} \sum_{r,s=1}^3 \sum_{m,n=-\infty}^\infty
a_{m(r)}^\dagger
\overline{N}^{(rs)}_{mn} a_{n(s)}^\dagger
\right] |0_{(1)}\rangle |0_{(2)}\rangle |0_{(3)}\rangle,
\end{equation}
where we have absorbed a constant into $f$ (it was undetermined anyway),
and we have introduced the matrices
\begin{equation}
\Gamma_{mn} = \sum_{r=1}^3 \sum_{p=-\infty}^\infty
\omega_{p(r)} X_{mp}^{(r)} X_{np}^{(r)},
\end{equation}
and
\begin{equation}
\overline{N}^{(rs)}_{mn} = \delta^{rs} \delta_{mn} -2 \sqrt{
\omega_{m(r)} \omega_{n(s)}} ( X^{(r) {\rm T}} \Gamma^{-1} X^{(s)})_{mn}.
\end{equation}

\subsection{Alternate, Simpler Derivation}

We now give a more straightforward way to arrive at the
same final result~(\ref{eq:vertexone}).
After Fourier transforming, the delta-functional can be expressed as
local conservation of momentum density on the worldsheet:
\begin{equation}
\Delta[ p_1(\sigma) + p_2(\sigma) + p_3(\sigma)].
\end{equation}
We are trying to find a
state $|V\rangle$ which is an oscillator representation
of the position- and momentum-space
delta-functionals.  Now recall the elementary identity
\begin{equation}
x \delta(x) = 0.
\end{equation}
The state $|V\rangle$ must therefore satisfy
\begin{equation}
\label{eq:laws}
\left(
p_1(\sigma) + p_2(\sigma) + p_3(\sigma) 
\right) 
|V\rangle
= 
(x_1(\sigma) + x_2(\sigma) - x_3(\sigma))|V\rangle
= 0.
\end{equation}
Let us take the Fourier transform of these equations with respect
to the $m$-th Fourier mode of string 3.
Then we make use of the same matrices $X^{(r)}$ introduced above, and
we find the following equations, which must vanish for each $m$:
\begin{eqnarray}
\label{eq:eqns}
\sum_{r=1}^3 \sum_{n=-\infty}^\infty X^{(r)}_{mn} p_{n(r)}|V\rangle = 0,
\qquad \sum_{r=1}^3
\sum_{n=-\infty}^\infty e(p^+_r) X^{(r)}_{mn} x_{n(r)}|V\rangle = 0.
\end{eqnarray}

If we make an ansatz for $|V\rangle$ of the form
\begin{equation}
|V \rangle  = f(p_1^+, p_2^+, p_3^+)
\exp
\left[ {1 \over 2} \sum_{r,s=1}^3 \sum_{m,n=-\infty}^\infty
a_{m(r)}^\dagger
\overline{N}^{(rs)}_{mn} a_{n(s)}^\dagger
\right] |0_{(1)}\rangle |0_{(2)}\rangle |0_{(3)}\rangle,
\end{equation}
for some coefficients $\overline{N}$,
and then expand the $x$'s and $p$'s appearing in~(\ref{eq:eqns}) into creation
and annihilation operators, then one obtains some matrix equations whose
unique solution is
\begin{equation}
\label{eq:neumann}
\overline{N}^{(rs)}_{mn} = \delta^{rs} \delta_{mn} -2 \sqrt{
\omega_{m(r)} \omega_{n(s)}} ( X^{(r) {\rm T}} \Gamma^{-1} X^{(s)})_{mn}.
\end{equation}
It is clear that $f(p_1^+, p_2^+, p_3^+)$ remains undetermined by this method.

\subsection{Summary}

We have written the Hamiltonian of string theory in light cone gauge
as a free term plus a cubic interaction.  
It turns out (at least for the bosonic
string) that this is the whole story!  
One can use this simple Hamiltonian
to calculate the string S-matrix, to arbitrary order in string
perturbation theory, with no conceptual difficulties.
(The remaining measure factor $f$ will be determined in the next lecture.)
Since this is a light-cone gauge quantization, the procedure is especially
simple.  There are no `vacuum' diagrams, so one just uses the simple Feynman
diagrammatic expansion
of the S-matrix:
the only interaction vertex is a simple string splitting or joining.

\section{Lecture 3: Light-Cone String Field Theory}

\subsection{Comments on the Neumann Coefficients}

In the last lecture we wrote the cubic string vertex as
a squeezed state in the three-string Fock space:
\begin{equation}
\label{eq:vertex}
|V\rangle = \delta(p_1^+ + p_2^+ + p_3^+) 
f(p_1^+, p_2^+, p_3^+,\mu) (\det \Gamma)^{(2-D)/2} \exp \left[
V(a_{(1)}^\dagger, a_{(2)}^\dagger, a_{(3)}^\dagger) \right] |0_{(1)}\rangle
|0_{(2)}\rangle |0_{(3)}\rangle. 
\end{equation}
Here $f(p_1^+, p_2^+, p_3^+,\mu)$ is a measure factor which we have not
yet determined and $D$ is the dimensionality of space time.  This enters
the formula because one gets one factor of $(\det \Gamma)^{-1/2}$ for each
dimension transverse to the light cone.  Finally we have made the convenient
definition
\begin{equation}
V(a_1, a_2, a_3) =
{1 \over 2}
\sum_{r,s=1}^3 \sum_{m,n=-\infty}^\infty \overline{N}^{(rs)}_{mn}
a_{m(r)} a_{n(s)}.
\end{equation}

The matrix element $\overline{N}^{(rs)}_{mn}$ expresses
the coupling between mode $m$ on string $r$ and mode $n$ on string $s$.
These coefficients are called Neumann coefficients.  Although the $X$ matrices
are independent of $\mu$, the matrix $\Gamma$ depends on $\mu$ (and
the three $p_+$'s) in a highly
nontrivial way.
In the $\mu \to 0$ limit, it is rather easy to show that these Neumann
coefficients reduce correctly to the flat space case, where explicit formulas
are known for $\overline{N}^{(rs)}_{mn}$ (see the papers
reproduced in \cite{Schwarz}).

A huge technology has been developed towards obtaining
explicit formulas for $\overline{N}^{(rs)}_{mn}$
as a function of $\mu$ and $p_+^r$.
This material is too technical to present in detail,
so we will just summarize the current state of the art \cite{He:2002zu}.
Recall that the dual BMN gauge theory is believed to be effectively
perturbative
in the parameter
\begin{equation}
\lambda' = {1 \over (\mu p^+ \alpha')^2}.
\end{equation}
So, in order to make contact with perturbative gauge theory calculations,
we are particularly interested in studying string interactions in the large $\mu$
limit.  In this limit it can be shown that
\begin{eqnarray}
&&\overline{N}^{(13)}_{mn} = {1 \over 2 \pi} { (-1)^{m+n+1} \sin(\pi n y) \over
p^+_3 \omega_{m(1)} + p^+_1 \omega_{n(3)} }
 \sqrt{y \over \omega_{m(1)} \omega_{n(3)} }
\cr
&&
\qquad\times\left[
\sqrt{(\omega_{m(1)} + \mu p^+_1 \alpha'
)(\omega_{n(3)} + \mu p^+_3 \alpha')}
+ \sqrt{(\omega_{m(1)} - \mu p^+_1 \alpha'
)(\omega_{n(3)} - \mu p^+_3 \alpha')}
\right]
\cr
&&\qquad\qquad+ {\cal{O}}(e^{-2 \pi \mu}, e^{-2 \pi \mu y}, e^{-2 \pi \mu(1-y)}).
\end{eqnarray}
The first term encodes all orders in a power series expansion in $\lambda'$.
Specifically,
\begin{equation}
\overline{N}^{(13)}_{mn} = \left[{ (-1)^{m+n+1} \over \pi \sqrt{y}}
{\sin(\pi n y)
\over n - m/y} + {\cal{O}}(\lambda') + \cdots \right]
+ {\mbox{nonperturbative}}.
\end{equation}
It is intriguing that the nonperturbative corrections look like
D-branes rather than instantons (i.e.~they are ${\cal{O}}(e^{-1/g})$ rather
than ${\cal{O}}(e^{-1/g^2})$).

We have only written the Neumann coefficient $\overline{N}^{(13)}$, but
in fact it is easily shown that in the $\mu \to \infty$ limit,
\begin{equation}
\label{eq:largemuN}
\overline{N}^{(13)}_{mn} = \sqrt{y} X^{(1) {\rm T}}_{mn},
\qquad
\overline{N}^{(23)}_{mn} = \sqrt{1-y} X^{(2) {\rm T}}_{mn},
\end{equation}
while  all other components are zero.
This fact actually has a very nice
interpretation in the BMN gauge theory, which we present in
Table~\ref{TABLE3}.

\begin{table}
\begin{center}
\begin{tabular}{|ccccc|} \hline
& & & & \cr
`three-point' functions
&  & $\longleftrightarrow$ & & matrix elements of $|V\rangle$
\cr
at $\lambda'=0$& & & & at $\mu=\infty$\cr
& $\nwarrow \searrow$ & & $\nearrow \swarrow$ & \cr
& & & &  \cr
& & matrix elements of $\Sigma$,& & \cr
& & splitting-joining operator& & \cr 
& & & & \cr \hline
\end{tabular}
\caption{Correspondence between the $\mu = \infty$ limit of
the three-string vertex, the BMN limit of the gauge theory,
and the splitting-joining operator $\Sigma$ in the
string bit model \cite{Vaman:2002ka}.
This correspondence has been explored in a number of papers, including
\cite{Chu:2002pd,Huang:2002wf,Pearson:2002zs,Chu:2003qd}.
\label{TABLE3}}
\end{center}
\end{table}

\subsection{The Consequence of Lorentz Invariance}

Our vertex~(\ref{eq:vertex}) still has an arbitrary function $f$ of the
light-cone momenta and $\mu$,
and a factor $(\det \Gamma)^{(2-D)/2}$, which is also a terribly
complicated function of the light-cone momenta and $\mu$.
In flat space ($\mu=0$), it was shown long ago that Lorentz-invariance of the vertex,
and in particular, the covariance of S-matrix elements under $J^{+-}$ Lorentz
transformations, requires $D=26$ and $f = (\det \Gamma)^{12}$.

This fact is nice for the oscillator representation since these
factors then cancel and~(\ref{eq:vertex}) 
can simply be written as
\begin{equation}
|V\rangle = \delta(p_1^+ + p_2^+ + p_3^+) 
\exp \left[
V(a_{(1)}^\dagger, a_{(2)}^\dagger, a_{(3)}^\dagger) \right] |0_{(1)}\rangle
|0_{(2)}\rangle |0_{(3)}\rangle,
\qquad {\rm for~}\mu=0,
\end{equation}
with no additional factors (except perhaps some innocent overall factors
like $2 \pi$'s which we have not carefully kept track of).

However, in the functional representation this fact is quite mysterious!
It means that the correct, Lorentz-invariant string vertex in flat space,
\begin{equation}
\label{eq:hardact}
H_3 = g_2 \int 
 \delta(p_1^+ + p_2^+ + p_3^+)
(\det \Gamma)^{12}
\Delta[ x_1(\sigma) + x_2(\sigma) - x_3(\sigma)]\prod_{r=1}^3
\left( dp_r^+ D x_r(\sigma)\ \Phi[ p_r^+, x_r(\sigma)]
\right),
\end{equation}
has a very peculiar measure factor $(\det \Gamma)^{12}$ which would have
been impossible to guess purely within the functional approach.
Moreover, since the function $(\det \Gamma)^{12}$
is a highly complicated function of the $p_i^+$,
if we Fourier transform the action~(\ref{eq:hardact})
back to $x^-$ position space, we find that it involves infinitely
many $\partial_{x^-}$ derivatives, in a very complicated way.
This tells us that light-cone string field theory is highly
non-local in the $x^-$ direction.

The plane wave background with $\mu>0$ has fewer bosonic symmetries
than flat space.  In particular, it
does not have the $J^{+-}$ or $J^{-I}$ symmetries.
This means that it is impossible to use Mandelstam's method to determine
what the corresponding 
measure factor is when $\mu>0$.  Our vertex for string interactions in the
plane wave
background remains ambiguous up to an overall (possibly very complicated)
function of $p_1^+, p_2^+, p_3^+$ and $\mu$. 

We determined the form of the vertex by requiring continuity of the string
worldsheet, but evidently that is not enough to solve our problem.
In the rest of this lecture we will learn why light-cone string field theory
works, and what the physics is that does
completely determine the light-cone
vertex (since continuity
is not enough).  To be precise, we should say that we will discuss the physics
which {\it in principle} determines the light-cone vertex uniquely.
The actual {\it calculation} of what this overall function is has not yet
been performed for $\mu>0$, and is likely rather difficult.
(In the supersymmetric theory, it has been speculated that supersymmetry
might fix this overall function essentially to 1, but this has not been
proven.)

The first step on this exciting journey into the details of
string field theory will be a close look at the four-string scattering
amplitude.

\subsection{A Four-String Amplitude}

We consider a $2 \to 2$ string scattering process at tree level.
This exercise will be useful for showing how to use the formalism
of light-cone string field theory to do actual calculations.
Without loss of generality we can choose to label the particles
so that 1 and 2 are incoming (positive $p^+$) and 3 and 4 are
outgoing (negative $p^+$) and furthermore
$-p_4^+ > p_1^+ > p_2^+ > -p_3^+$.

\begin{figure}[htb]
\centerline{\hbox{ \psfig{figure=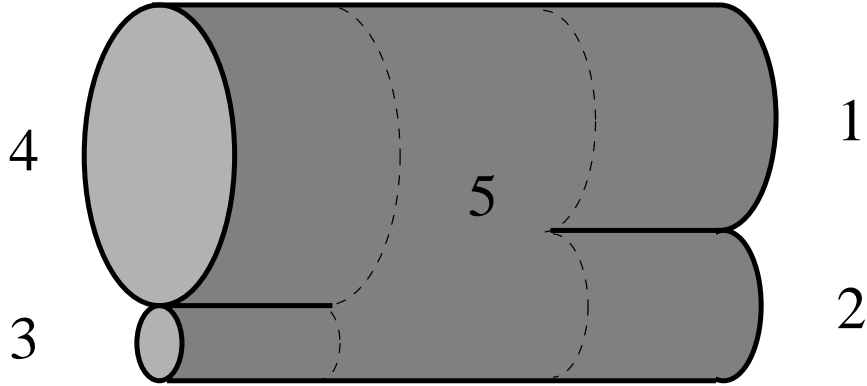} }}
\caption{\footnotesize The $s$-channel contribution
to the  tree level 4-particle
interaction $1 + 2 \to 5 \to 3 + 4$.  \label{fig:FIG2}}
\end{figure}

The $s$-channel amplitude (see Figure~\ref{fig:FIG2}) is
\begin{eqnarray}
&&{\cal{A}}_s=
\int_0^\infty dT_5 \int_0^{p^+_5} {d \sigma_5
\over p^+_5}
\underbrace{\langle 0_{(5)} | {V(a^{}_4,a_3^\dagger,a_4^\dagger)}
|0_{(3)} \rangle |0_{(4)}\rangle}_{5 \to 3 + 4}
\cr
&&\qquad\qquad\qquad\times
{e^{-T_5 (E_1 + E_2 - H_{(5)}) + 2 \pi i \sigma_5
(N_{(5)}- \widetilde{N}_{(5)})/p^+_5}}
\underbrace{\langle 0_{(1)} | \langle 0_{(2)}| {V(a^{}_1, a^{}_2, a_5^\dagger)
} |0_{(5)}\rangle}_{1 + 2 \to 5}.
\end{eqnarray}
Let us explain each ingredient.  First of all, trivial overall $p^+$-momentum conserving
delta functions are always understood but have not been written in order
to save space.  The processes $1 \to 2 + 5$ and $5 \to 3 + 4$ are as
indicated, making use of our vertex function $V$.
In between these two we have inserted the light-cone
propagator for the intermediate
string 5:
\begin{equation}
\label{eq:propagator}
{1 \over E - H} = \int_0^\infty dT e^{-T(E-H)}.
\end{equation}
By $H_{(5)}$ we mean of course the Hamiltonian for string 5:
\begin{equation}
H_{(5)} = {1 \over p^+_{(5)}} \sum_{n=-\infty}^\infty
\omega_{n(5)} a_{n(5)}^\dagger a_{n(5)}.
\end{equation}
Finally, the integral over $\sigma_5$ enforces the physical state condition
on the intermediate string by projecting onto those states which satisfy
\begin{equation}
N_{(5)} - \widetilde{N}_{(5)} = \sum_{n=-\infty}^\infty n a_{n(5)}^\dagger a_{n(5)}
= 0.
\end{equation}

The full amplitude has two additional contributions.
In the $t$-channel, we have first $1 \to 3 + 6$, and then $6 + 2 \to 4$:
\begin{eqnarray}
&&{\cal{A}}_t=
\int_{-\infty}^0 dT_6 \int_0^{p^+_6}
{d \sigma_6 \over p^+_6}
\langle 0_{2,6} | {V(a^{}_2, a^{}_6, a_4^\dagger)}
|0_4\rangle
\cr
&&\qquad\qquad\qquad\times
e^{-T_6(E_1 - E_3 - H_{(6)}) +2 \pi i \sigma_6
( N_{(6)} - \widetilde{N}_{(6)})/p^+_6}
\langle 0_1 | {V(a^{}_1, a_3^\dagger, a_6^\dagger)
} |0_{3,6} \rangle.
\end{eqnarray}
Finally in the $u$-channel, $2 \to 3 + 7$ and then $7 + 1 \to 4$:
\begin{eqnarray}
&&{\cal{A}}_u = \int_{-\infty}^0 dT_7
\int_0^{p^+_7} { d\sigma_7 \over
p^+_7} \langle 0_{1,7} | {V(a^{}_1, a^{}_7, a_4^\dagger)}
|0_4\rangle
\cr
&&\qquad\qquad\qquad\times
e^{-T_7(E_2 - E_3 - H_{(7)}) +2 \pi i \sigma_7
( N_{(7)} - \widetilde{N}_{(7)})/p^+_7}
\langle 0_2 | {V(a^{}_2, a_3^\dagger, a_7^\dagger)
} |0_{3,7} \rangle.
\end{eqnarray}

What are these things?  Well, each ${\cal{A}}$ is just a state
in ${\cal{F}}^4$, the fourth power of the string Fock space.
If we want to know the amplitude for scattering four particular external
states, then we just have to calculate
\begin{equation}
\langle 3 | \langle 4| {\cal{A}} |1 \rangle | 2\rangle
\end{equation}
(summed over channels) to get the scattering amplitude
as a function of $p^+_i$ and $\mu$.

It should be emphasized that the harmonic oscillator algebra
gives very complicated functions of $T$ and $\sigma$ which need
to be integrated over.  Actually performing this calculation is
far outside the scope of these lectures (see
\cite{Cremmer:1974ej}), but we would like to make one
very important point about the general structure of this amplitude.

\begin{figure}[htb]
\centerline{\hbox{ \psfig{figure=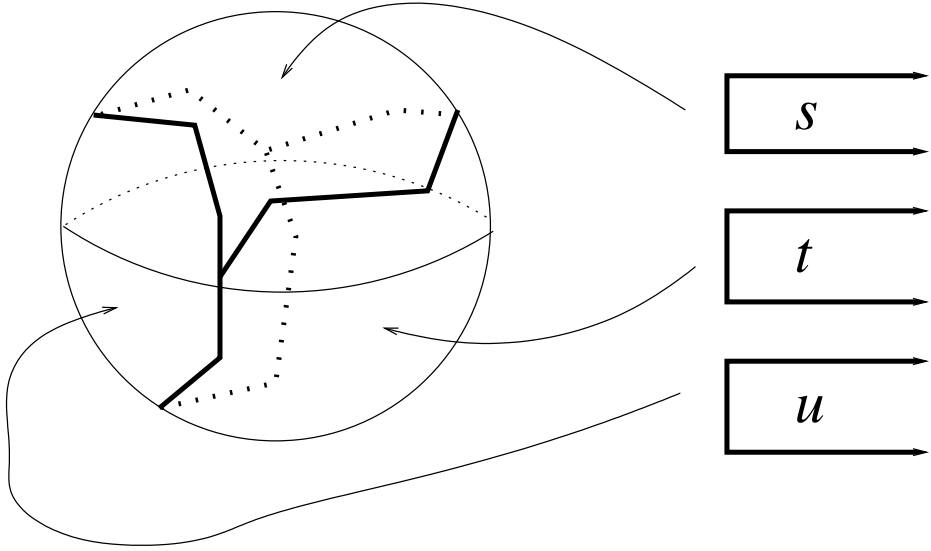} }}
\caption{\footnotesize A schematic picture of the moduli map
for the tree level 4-particle interaction.
\label{fig:FIG3}}
\end{figure}

Let us denote ${\cal{R}}_i = \{ (T_i, \sigma_i):
T \in [0,\infty), \sigma_i \in [0,p^+_i] \}$, which are the
two dimensional regions over which the quantities ${\cal{A}}_i$
must be integrated.  There exists a particular
map\footnote{We are 
not aware that any name has been given to this map in the literature.
We will call it the `moduli map.'  It should not be confused with a
very much related Mandelstam map
from a light-cone diagram with {\bf fixed}
moduli into the complex plane.}
$z(T_i, \sigma_i)$
which patches together
these three coordinate regions onto a sphere as shown in Figure~\ref{fig:FIG3}.
This much is of course obvious.

Let us define ${\cal{A}}(z)$ to be the image of the three individual
${\cal{A}}_s$, ${\cal{A}}_t$, and ${\cal{A}}_u$ on the sphere, patched
together via the moduli map.
It turns out that in flat space, precisely in the critical dimension
$D=26$, the function ${\cal{A}}(z)$ on the sphere is continuous along
the boundaries between the images of ${\cal{R}}_i$ (that is, continuous
along the dark lines in Figure~\ref{fig:FIG3}),
which means that the amplitude can be written as
\begin{equation}
\label{eq:moduli}
\sum_{i=s,t,u} \int_{{\cal{R}}_i} dT_i d
\sigma_i  {\cal{A}}_i  = \int_{S^2} d^2 z \ {\cal{A}}(z).
\end{equation}

\subsection{Why Light-Cone String Field Theory Works}

The right hand side of~(\ref{eq:moduli}) has a very familiar form.
When we studied string theory, we learned that in order to calculate
a four-string amplitude at tree level in closed string theory, one inserts
four vertex operators on the sphere.  The positions of three vertex operators
can be fixed using the conformal Killing vectors, and one is left with
some amplitude (depending on the particular vertex operators inserted) which
must be integrated over $z$, the position of the remaining vertex operator.
The moduli space of a sphere with four marked points (the positions of vertex
operators) is therefore
the sphere itself.

It turns out that the integrand ${\cal{A}}(z)$ on the right-hand side
of~(\ref{eq:moduli}) is precisely the integrand one would derive
from the Polyakov path integral in the covariant formulation of string theory.
Although we have studied only one of the most trivial possible
amplitudes, the equation~(\ref{eq:moduli}) indicates a very general
feature.  Amplitudes calculated in light-cone string field theory,
with any number of external states and at arbitrary order in
string perturbation theory,
are precisely equivalent to those calculated using the covariant
Polyakov path integral \cite{D'Hoker:1987pr}.
This equivalence relies
on two important facts:

\noindent
{\bf Property 1: Triangulation of Moduli Space.}
 Consider all of the light-cone diagrams which contribute to an
amplitude with $g$ closed string loops and $n$ external particles
\cite{Giddings:1986rf}.
The diagrams will be labeled by $6g+2n-6$ parameters:  $g$
`$p^+$-momentum fractions', $3g+n-3$ twist angles
(to impose the physical state condition on
intermediate string states), and $2g+n-3$ interaction
times.
The first important fact is that the moduli map provides a one-to-one
map between this $6g+2n-2$-dimensional parameter space and the moduli space of
Riemann surfaces of genus $g$ with $n$ marked points (the locations of the
vertex operators).  A mathematical way of saying this is that the
light-cone vertex provides a triangulation of the moduli space ${\cal{M}}_{g,n}$.

\noindent
{\bf Property 2: The Measure on Moduli Space.}
The second important fact is that the integrand of the light-cone
vertex, including all of the complicated structure involving the Neumann
matrices and determinants thereof, maps under the moduli map to precisely
the correct integration measure which arises from the Polyakov path integral!

The proof of these remarkable facts would take us too far afield, but
we cannot stress enough the importance of these facts, which are deeply
rooted in the underlying beautiful consistency of string theory.
In fact, this equivalence can be used to prove the unitarity of 
the Polyakov path integral \cite{D'Hoker:1987pr}:
although the path integral is not manifestly
unitary,
it is equivalent to the light-cone formalism, which is
manifestly unitary!

We are now in a position to answer some questions which may have been bothering
some students since the last lecture:  why is it sufficient to consider a
cubic interaction between the string fields, and why
is it sufficient to consider the simplest possible cubic interaction,
with only a delta-functional (and, for example,
no derivative terms like $\delta \Phi[x(\sigma)]/\delta
x(\sigma_{\rm I})$)?  The answer is that
the simple
cubic interaction is sufficient because (1) the iterated cubic interaction
covers precisely one copy of moduli space and (2) the vertex we wrote
down precisely reproduces the correct integration measure on this moduli space.
We don't need anything else!

It is sometimes said that the symmetry
algebra (in particular, the supersymmetry algebra, for superstrings),
uniquely determines the interacting string Hamiltonian to all orders
in the string coupling.  This is a little bit misleading.
For example,
in the supersymmetric theory one could take $Q = ({\rm anything})$ and
then define $H = ({\rm anything})^2$, and as long as (anything)
commutes with rotations and translations, one would have a realization
of the symmetry algebra!
The symmetry argument, however, provides
no motivation for considering only a cubic interaction.
The true criteria are (1) and (2) listed above, and fortunately it
turns out to be true that (at least for the bosonic string), one can
find a purely cubic action with properties (1) and (2).  

\subsection{Contact Terms}

Now, fact number (1), that the cubic delta-functional vertex covers
moduli space precisely once, is essentially a mathematical theorem about
a particular cell decomposition of ${\cal{M}}_{g,n}$ that holds
quite generally \cite{Giddings:1986rf}.
However, (2) can fail in subtle ways in certain circumstances.

In particular, it can happen that one or more of the ${\cal{A}}_i$'s
has singularities in moduli space.
A typical case might be for example that ${\cal{A}}_i(T,\sigma) \sim
{1 \over T^3}$ near $T=0$, which is not integrable.
This gives rise to divergences in string amplitudes, which need
to be corrected by adding new string interactions
to the Hamiltonian.  However, these interaction terms are always delta-function
supported on sets of measure zero ($T=0$ in this example) in moduli space,
and therefore they do not spoil the beautiful triangulation that the cubic
vertex provides.
As long as we don't add any interaction with {\it finite} measure, the
triangulation still works just fine.

\bigskip
\noindent
{\bf Definition.}  We define a {\bf contact term} to be any 
term in the Hamiltonian which has support only on a set of measure zero
in moduli space.

\bigskip
\noindent
{\bf Corollary.}  All contact terms are divergent.
Proof: If they were finite,
they wouldn't give any contribution, there would be no point
to include them, since by definition they are integrated over
sets of measure zero!

\bigskip

In flat space, it is known \cite{D'Hoker:1987pr}
that the bosonic string requires no contact
terms, while the IIB superstring is widely (though not universally)
believed to require an infinite
number of contact terms.  The word `believe' can be thought of in the following
sense:  since the purpose of contact terms is to eliminate divergences (and
indeed we will see how they arise from short-distance singularities on the
worldsheet), one can
think of a contact term as a counterterm in the sense of renormalization.
Now, there are infinitely many such counter terms that one can write down for the
IIB string,
and while some of them may have coefficients which are equal
to zero, it is widely believed that infinitely many of them will have nonzero
coefficients.

For example, in open superstring field theory, it was argued
in \cite{Green:fu} that a possible contact term in the $s$-channel
of the $2 \to 2$ amplitude in fact vanishes.  However, it was argued
in \cite{Greensite:1986gv} that there is a contact term for this
process in the $u$-channel.  For higher amplitudes the situation is
much more complicated and has not been addressed in detail.
A non-zero contact term in the one-loop mass renormalization has been
studied in the plane wave background \cite{Roiban:2002xr} and will
be discussed in the next lecture.

For strings in the plane wave background, the question
of whether property (2) holds has not been addressed,
mostly because we do not have the analogue of the covariant Polyakov formalism
in which we can actually calculate anything.  First we would
need to calculate
this overall factor $f(p_1^+, p_2^+, p_3^+, \mu)$ and then see if there
are any divergences which give rise to contact terms.

Any of the contact terms in IIB string theory in flat space will surely
give rise to $\mu$-dependent contact terms in the plane wave background.
In principle there could be new contact terms introduced which go to
zero in the limit $\mu \to 0$.  Certainly we do not know how to disprove
such a possibility, but we believe this is unlikely:  contact terms may be
thought of as coming from short-distance singularities on the string
worldsheet, but the addition of a mass parameter $\mu$ on the worldsheet
should not affect any of the short-distance behavior.

In fact, it is more likely that the opposite
is true: that there are infinitely many contact terms in flat space,
but all but a finite number vanish in the plane wave background
when $\mu$ is large \cite{Pearson:2002zs}.
We will have more to say about this in the next
lecture.

\subsection{Superstrings}

Let's go back to the beginning of
Lecture~2, but add fermions to the picture.
We consider now a superparticle on the plane wave solution of IIB supergravity.
The physical 
degrees of freedom of the theory are encoded in a superfield $\Phi(x,\theta)$
which has an expansion of the form
\cite{Green:1982tk,Metsaev:2002re}
\begin{equation}
\Phi(p^+, x,\theta) = (p^+)^2 A(x) + p^+ \theta^a \psi_a(x) +
\theta^{a_1} \theta^{a_2} p^+ A_{a_1 a_2}(x) + \cdots + {1 \over (p^+)^2}
\theta^8 A^*(x),
\end{equation}
where $\theta$ is an eight-component SO(8) spinor
($\theta^8$ is short for eight powers of $\theta$ contracted with
the fully antisymmetric tensor $\epsilon$).
Initially we allow all the component fields to be complex, but this gives
too many components (256 bosonic $+$ 256 fermionic) so we impose
the reality condition
\begin{equation}
\label{eq:superfield}
\Phi(x, \theta) = (p^+)^4 \int d^8 \theta^\dagger e^{i \theta \theta^\dagger/p^+}
(\Phi(x,\theta))^\dagger
\end{equation}
which cuts the number of components in half.
Note, in particular, that this constraint correctly gives the
self-duality condition for the five-form field strength.

When we second quantize, this hermiticity condition implies that
the inner product on the string field theory Hilbert space ${\cal{H}}$
is {\bf not} the inner product na\"ively inherited from the single
string Hilbert space.  Instead,
\begin{equation}
\label{eq:hermiticity}
{\cal{A}}_{|a \rangle}(p^+)^\dagger = {\cal{A}}_{|a'\rangle}(-p^+),
\end{equation}
where the states $|a\rangle$ and $|a'\rangle$ differ by reversing
the occupation of all of the fermionic zero modes, i.e.
if $|a\rangle = |0\rangle$, then $|a'\rangle = \theta^8 |0\rangle$, etc.

The action for the free superparticle 
is
\begin{equation}
\label{eq:superaction}
S = {1 \over 2} 
\int d^{10} x d^8 \theta
\ \Phi ( \nabla^2 - 2 i \mu \partial_- \theta \Pi \partial_\theta) \Phi,
\end{equation}
where $\Pi = \Gamma^1 \Gamma^2 \Gamma^3 \Gamma^4$.
The quantity in brackets is the quadratic Casimir of the plane wave superalgebra.
It is straightforward to insert the superfield~(\ref{eq:superfield})
into~(\ref{eq:superaction}) and find the resulting spectrum
\cite{Metsaev:2002re}.

The action~(\ref{eq:superaction})
of course may also be obtained simply by linearizing the action for
IIB supergravity around the plane wave background, and the spectrum may
be obtained by linearizing the equations of motion around the background
and finding the eigenmodes.
This has been worked out in detail in \cite{Metsaev:2002re}, but we will use
only one fact which emerges from this analysis.
It turns out that there is a unique state
with zero energy, which we will call $|0\rangle$.
The corresponding spacetime field is a linear combination of 
the trace of the graviton over four of the eight transverse
dimensions, $h_{ii}$,
and the components of the four-form gauge potential $a_{1234}$ in the first
four directions.
This field lives in the $\theta_{\rm R}^4$ component of the
superfield, where we define left and right chirality with respect
to $\Pi$ (i.e., $\theta_{\rm R,L} = {1 \over 2} (1 \pm \Pi) \theta$).
The only important fact which you might want to keep
in mind is that this spacetime field is {\bf odd} under the ${\mathbb{Z}}_2$
symmetry
which exchanges the two SO(4)'s:
\begin{equation}
\label{eq:ztwo}
Z{ \cal{A}}_{|0\rangle} = - {\cal{A}}_{|0\rangle} Z.
\end{equation}
The full string theory Hamiltonian (including interactions)
commutes with the ${\mathbb{Z}}_2$ operator $Z$, and this fact
together with~(\ref{eq:ztwo}) can be used to derive useful
selection rules for string amplitudes.

When we promote the superfield to string theory, it becomes a functional
of the embedding of the string into superspace:
$\Phi[p^+, x(
\sigma), \theta(\sigma)]$.  The cubic interaction term
has a delta-functional for continuity of $x(\sigma)$, and also
a delta-functional for the superspace coordinates:
\begin{equation}
\Delta[ \theta_1(\sigma) +\theta_2(\sigma) - \theta_3(\sigma)].
\end{equation}
One can write this delta-functional in an oscillator representation as
a squeezed state involving the fermionic creation operators.
The `fermionic' Neumann matrices are easily obtained from
the bosonic Neumann matrices.

\section{Lecture 4: Loose Ends}

\subsection{The `Prefactor'}

 From the preceding section one might have the impression that
light-cone string field theory for fermionic strings is a relatively
trivial modification of the bosonic theory.  Unfortunately, this is
not true.  In the fermionic theory the cubic  string interaction
is no longer a simple $\Phi^3$ vertex with derivatives.
Instead it is quadratic in
string field functional derivatives.  One way to see why this is
necessary is to look at the supersymmetry algebra, which constrains
the form of the Hamiltonian and dynamical supercharges.

The (relevant part of the) spacetime supersymmetry algebra is
\begin{equation}
\label{eq:susy}
\{ Q^-, \overline{Q}{}^- \} = 2 H, \qquad [ Q^-, H] = 0, \qquad
[ \overline{Q}{}^-, H ] = 0.
\end{equation}
At the free level, a realization of this algebra is given by the free
Hamiltonian $H_2$ we met before, and the free
supercharges $Q^-_2$, which are given by
\begin{equation}
Q_2^- = \int dp^+ \ p^+ \Phi^\dagger q^- \Phi,
\end{equation}
where the worldsheet supercharge is
\begin{equation}
\label{eq:wssc}
q^- = \int d\sigma\left[ 4 \pi e(p^+_r) p^I \gamma_I \lambda - {i \over 4 \pi}
\partial_\sigma x^I \gamma_I \theta - i \mu x^I \gamma \Pi \lambda \right].
\end{equation}
Here $\lambda$ is the `fermionic momentum' conjugate to
$\theta$ (i.e., it is just $\overline{\theta})$).

When we turn on an interaction $H_3$ in the Hamiltonian,
we also need to turn on interactions $Q_3$, $\overline{Q}{}_3$
in the dynamical supercharges
to ensure that
the generators
\begin{equation}
H = H_2 + g_s H_3, \qquad
Q^- = Q^-_2 + g_2 Q^-_3, \qquad
\overline{Q}^- = \overline{Q}^-_2 + g_s \overline{Q}^-_3
\end{equation}
provide a (non-linear) realization of the supersymmetry
algebra~(\ref{eq:susy}).

Now there is a simple argument (see chapter 11 of \cite{GSW})
which shows that the choice
$|H_3\rangle = |V\rangle$ would be incompatible with the supersymmetry
algebra.  Consider
the relation $0=[ \overline{Q}{}^-, H]$ at first order in the string coupling.
This gives
(via the state-operator correspondence)
\begin{equation}
\label{eq:check1}
0 =  \sum_{r=1}^3 H_{(r)} |Q^-\rangle + \sum_{r=1}^3 Q_{(r)}^- |V\rangle.
\end{equation}
Now let us consider (for example) a matrix element of this relation
where we sandwich three on-shell states on the left.
Then $\sum_{r=1}^3 H_{(r)}$ acts to the left and gives zero, leaving us
with only the second term.
Now the state $|V\rangle$ indeed is annihilated by the constraints
\begin{equation}
\label{eq:conser}
\sum p_{(r)}|V\rangle=  \sum \lambda_{(r)}
|V\rangle= \sum e(p^+_r) x_{(r)}|V\rangle=
\sum e(p^+_r) \theta_{(r)}|V\rangle = 0.
\end{equation}
Now after looking at~(\ref{eq:wssc}),
the conditions~(\ref{eq:conser})
seem to imply that $Q_{(r)}^- | V\rangle = 0$, and hence
that the desired relation~(\ref{eq:check1}) is true.

However, one can check that the operators in~(\ref{eq:wssc}) are
actually singular near the interaction point.  For example, we have
$p(\sigma) \lambda(\sigma)|V\rangle \sim \partial_\sigma
x(\sigma) \theta(\sigma)
|V\rangle \sim \epsilon^{-1} |V\rangle$ near $\sigma = \sigma_{\rm I}$.
Therefore, although $\sum_{r=1}^3 Q_{(r)}^-|V\rangle$
vanishes pointwise in $\sigma$ (except at $\sigma=\sigma_{\rm I}$), the singular
operators nevertheless give a finite contribution when integrated over $\sigma$.
This contribution can be calculated by deforming
the $\sigma$ contour in an
appropriate way and reading off
the residue of the pole at $\sigma = \sigma_{\rm I}$.

By calculating the residue of this pole, it can be shown that in order to
supersymmetrize the vertex, it is necessary to introduce some
operators (called `prefactors')
$\hat{h}$, $\hat{q}{}^-$, $\overline{\hat{q}}{}^-$
such that the interacting Hamiltonian and supercharges are given by
\begin{equation}
|H_3\rangle = \hat{h} |V\rangle, \qquad
|Q_3\rangle = \hat{q} |V\rangle, \qquad
|\overline{Q}{}_3\rangle = \overline{\hat{q}}{}^- |V\rangle.
\end{equation}
It turns out that $
\hat{h}$ is a second-order polynomial in bosonic mode-creation
operators (the $a^\dagger$'s) while $\hat{q}$
and $\overline{\hat{q}}{}^-$ are linear in bosonic creation operators.
They also have a very complicated expansion in terms of fermionic
modes, and we will not give the complete formula here.

It is essential to note, however, that the last term in~(\ref{eq:wssc}) is
non-singular when acting on $|V\rangle$.
This makes sense, since the parameter $\mu$ introduces a scale
in the worldsheet theory, but this should not affect the short
distance physics.  Therefore the functional form of the prefactor
has essentially the same form as in flat space (there are subtleties
in passing from the functional representation to the oscillator representation,
though).

We have shown that in the functional representation,
the cubic interaction between three string super-fields is {\bf not}
given simply
by the delta-functionals
\begin{equation}
\Delta[x_1(\sigma) + x_2(\sigma)  - x_3(\sigma)] \Delta[ \theta_1(\sigma)
+ \theta_2(\sigma) - \theta_3(\sigma)] \prod_{r=1}^3 \Phi[ x_r(\sigma),
\theta_r(\sigma)]
\end{equation}
In addition, there is a complicated combination of functional derivatives
acting on the $\Phi$ fields, inserted at the point $\sigma_{\rm I}$ where the
strings split.  This interaction point operator is sometimes
called the `prefactor'.
The presence of this prefactor is associated with the picture changing
operator in the covariant formulation.

\subsection{Contact Terms from the Interaction Point Operator}

\begin{figure}[htb]
\centerline{\hbox{ \psfig{figure=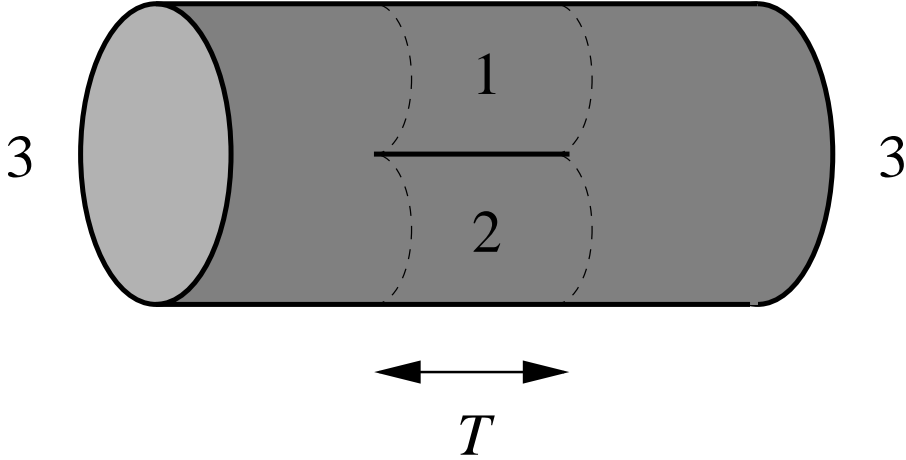} }}
\caption{\footnotesize The one-loop mass renormalization of string 3.
\label{fig:FIG4}}
\end{figure}

The prefactor $\hat{h}$ is an operator of weight ${3 \over 2}$, which 
means that at short distances we have $\hat{h}(x) \hat{h}(y) \sim (x-y)^{-3}$.
(In light-cone gauge we don't have a conformal
field theory in the pp-wave, so by `weight' we simply mean the
strength of the coincident singularity.)
A light-cone string diagram in which two (or more) of these prefactors
come very close to each other will therefore be divergent.  The simplest
example occurs in the two-particle amplitude at one loop (i.e., a contribution
to the one-loop mass renormalization), shown in Figure~\ref{fig:FIG4}.
This amplitude has been studied in the $\mu \to \infty$ limit of
the plane wave background in \cite{Roiban:2002xr}.

This amplitude has an integral $\int_0^\infty dT$ over the Schwinger
parameter giving the light-cone time between the splitting and joining
(think of it as coming from the propagator~(\ref{eq:propagator})
for the intermediate state),
but the integrand is divergent like $T^{-3}$ due to the colliding prefactors.
It is clear that at higher order in the string coupling (and/or with more
external states), we can draw diagrams which have arbitrarily many
colliding prefactors.  These divergent contributions to string amplitudes
must be rendered finite by the introduction of (divergent) contact interactions
as discussed in the previous lecture.  The belief is that there is a unique
set of contact interactions which preserves all the symmetries (Lorentz
invariance, supersymmetry) and which renders all amplitudes finite.
But these contact terms are very unwieldy, and almost impossible to calculate
explicitly, so they haven't really been studied in very much detail.

We will have a little bit more to say about these contact terms below.

\subsection{The $S$-matrix in the BMN Correspondence}

In past couple
of lectures we have demonstrated how to determine the light-cone Hamiltonian
of second-quantized IIB string theory in the plane wave background.  The two
remaining ambiguities are (1) that we have not determined some overall
factor $f(p_1^+, p_2^+, p_3^+, \mu)$ that appears in the cubic coupling,
and (2) that there are (probably) infinitely many contact counterterms
which need to be added to the action.
In principle (although probably not in practice),
the light-cone string field theory approach allows
one to calculate
 $S$-matrix elements to arbitrary order in the string coupling
via a quite straightforward Hamiltonian approach:  there is a (large)
Hilbert space of states, with a light-cone Hamiltonian acting on it, and
one can easily apply the rules of quantum mechanical perturbation theory
to give the $S$-matrix
\begin{equation}
\langle 1 | S |2 \rangle = \langle 1 | 1 -  2 \pi i \delta(E_1 - E_2) T(E + i \epsilon)
|2\rangle,
\end{equation}
where $H_2 |E_i\rangle = E_i$ and $T(z)$ is the transition operator
\begin{equation}
T(z) = V + V G(z) V, \qquad V \equiv H - H_2, \qquad G(z) = (z - H)^{-1}
\end{equation}
which is usually calculated via the Born series
\begin{equation}
T(z) = V + V G_0(z) V + V G_0(z) V G_0(z)V + \cdots,
\end{equation}
where $G_0(z) = (z- H_2)^{-1}$ is the `bare' propagator.

Typically, the $S$-matrix is the only good `observable' of string theory.
Local observables are not allowed because string theory
is a theory of quantum gravity, and in particular has diffeomorphism invariance.
The question is then, how is the string $S$-matrix encoded in the BMN limit
of ${\cal{N}} = 4 $ Yang-Mills theory?
Note that we are not talking about the $S$-matrix of the gauge theory
(which doesn't exist, since it is a conformal
field theory).  Instead, the question is about
how the string theory $S$-matrix can be extracted from
the BMN limit of the gauge theory.

\subsection{The Quantum Mechanics of BMN Operators}

It has been emphasized by a number of authors
that it can be useful
to think of gauge theory in the BMN limit as a quantum mechanical system
\cite{Gross:2002mh,Vaman:2002ka,Beisert:2002ff}.
There is a space of states
(the BMN operators), an inner product (the free
gauge theory two-point function),
and a Hamiltonian, given by $\Delta - J$.  Perturbation theory in the
gauge theory organizes itself into the two parameters
\begin{equation}
\lambda' = {\lambda \over J^2} = {g_{\rm YM}^2 N \over J^2}, \qquad
g_2 = {J^2 \over N},
\end{equation}
which are respectively the effective 't Hooft coupling and the effective
genus counting parameter, respectively.

Recall that the spectrum of BMN operators takes the following form
\begin{equation}
\label{eq:spec}
\Delta - J =\sum_{n=-\infty}^\infty N_n \sqrt{1 + \lambda' n^2},
\end{equation}
where $N_n$ is the number of impurities with phase $n$.
To one-loop, the gauge theory Hamiltonian $H = \Delta - J$ takes the
following form \cite{Beisert:2002ff}
\begin{equation}
\label{eq:gtH}
H = \underbrace{\sum_{n =- \infty}^\infty N_n + \lambda' \sum_{n=-\infty}^\infty
{1 \over 2} n^2 N_n}_{H_0} + \underbrace{\lambda' g_2 (H_+ + H_-)}_{ V},
\end{equation}
where $H_+$ and $H_-$ respectively increase and decrease the number of traces.
That is, if we act with $H_+$ on a $k$-trace operator, then it `splits' one
of the traces so that we get a $k+1$-trace operator.  Similarly $H_-$ `joins'
two traces.

The first two terms in~(\ref{eq:gtH}) are 
clearly just the first two terms in~(\ref{eq:spec}), expanded to order $\lambda'$,
so we have labeled them $H_0$ --- they constitute the `free' Hamiltonian.
The third term in~(\ref{eq:gtH}) has the structure of a three-string vertex,
and incorporates the string interactions, so we have labeled this term $V$.

The next step is to recall that we should keep in mind
the basis transformation that appeared in the lectures of H.~Verlinde
in this school.
In our first lecture we identified a precise correspondence between
single-trace operators
in the gauge theory and states in string theory.  It is natural therefore to
identify a double-trace operator in the gauge theory with the corresponding
two-particle state in the string theory, etc.
However this identification breaks down at $g_2 \ne 0$.
One way to see this is to
note that in string theory, a $k$ particle state and an $l$ particle state
are necessarily orthogonal (by construction)
for $k \ne l$.  However in the gauge theory,
it is not hard to check that the gauge theory overlap (given by the two-point
function in the free theory) is typically given by
\begin{equation}
\langle O_{\rm k-trace}(x) O_{\rm l-trace}(0) \rangle\sim
g_2^{|k-l|}.
\end{equation}

It has been conjectured \cite{Vaman:2002ka} that one
can write an exact formula, valid to all orders in $g_2$, for the 
inner product:
\begin{equation}
\label{eq:ip}
\langle 1 | 2 \rangle = \left(e^{g_2 \Sigma}\right)_{12},
\end{equation}
where $\Sigma$ is  the simple `splitting-joining'
operator of the bit model (see Table~\ref{TABLE3}).
The inner product~(\ref{eq:ip}) is diagonalized by the basis transformation
$S^{1/2} = e^{g_2 \Sigma/2}$.
Therefore, we propose the following BMN identification at finite $g_2$
\cite{Pearson:2002zs}:
\begin{eqnarray}
|0; p^+\rangle &\Longleftrightarrow& S^{-1/2} {\rm Tr}[Z^J],\cr
(a^i_0)^\dagger |0; p^+\rangle &\Longleftrightarrow& S^{-1/2} {\rm Tr}[\phi^i Z^J],\cr
(a^i_n a^j_{-n})^\dagger |0; p^+\rangle &\Longleftrightarrow& S^{-1/2}
\sum_{k=0}^J
e^{2\pi i k n/J}
{\rm Tr}
[ \phi^i Z^k \phi^j Z^{J-k}], \cr
&{\mbox etc.}&
\end{eqnarray}
Now the multi-particle states constructed from the operators on the right
hand side of this correspondence will have the property that $k$-string states
are orthogonal to $l$-string states for $k \ne l$.
However, we have lost the identification of `number of traces' with `number of
strings'.  Instead, we have something of the form
\begin{eqnarray}
&&{\mbox{$k$-string state}} \Longleftrightarrow [{\mbox{$k$-trace
operator}}]\cr
&&\qquad\qquad + g_2 \times
[ {\mbox{$k{-}1$-trace operator}} + {\mbox{$k{+}1$-trace operator}} ] + \cdots
\end{eqnarray}

It is convenient to perform this basis transformation on the operator $H$,
to define what we will call the string Hamiltonian $\widetilde{H}$, given by
\begin{equation}
\widetilde{H} = S^{1/2} H S^{-1/2} \equiv H_0 + W,
\end{equation}
for some new interaction $W$ (which is easily calculated).
Now $\widetilde{H}$ is simply a non-relativistic quantum
mechanical Hamiltonian,
and it is straightforward to derive from it an $S$-matrix.  This $S$-matrix
should be that of IIB string theory on the plane wave background.
It is important to recognize that this $S$-matrix is not unitarily
equivalent to the $S$-matrix obtained from the Hamiltonian $H_0 + V$
\cite{Spradlin:2003bw}.

\subsection{Contact Terms from Gauge Theory}

Previously we compared contact terms to counterterms, and
we saw how certain contact terms
in the superstring come about from regulating singularities that arise
when two operators collide on the worldsheet.  One could imagine
regulating the theory in some way in  order to render these
divergences finite.  One natural regulator, which is suggested by both
the dual gauge theory and the string bit model, is to discretize the
worldsheet.

Discretizing the worldsheet leads to a spacetime Hamiltonian
which depends on $J$, the total number of bits.  Schematically, we might
have something like
\begin{equation}
\label{eq:contact}
H(J) = H_2(J) + H_3(J) + \sum_{k=1}^\infty J^k C_k(J),
\end{equation}
where we suppose that $C_k(J)$ are finite in the
$J \to \infty$ limit.
When $J$ is finite, the cubic interaction $H_3(J)$ will fail to
precisely cover the moduli space, but the additional
`contact terms' will be finite
and will cover regions of moduli space of small but nonzero measure.
In the continuum limit $J \to \infty$, the contact terms become infinite
but restricted to regions of measure zero.

Physical quantities (such as $S$-matrix elements) should of course be
independent of the regulator $J$.  The precise way to say this is
that the amplitudes obtained from $H(J_1)$ differ from those obtained
from $H(J_2)$, $J_1 \ne J_2$ by something which is BRST-exact.
Anything BRST-exact
integrates to zero over moduli space, so this is a sufficient
condition for amplitudes to indeed be independent of the cutoff $J$.

The previous two paragraphs have been well-motivated, but not precise:
in fact, we do not know how to make precise
sense of $H(J)$ in string theory.
Certainly discretized string theories have been considered in the past
(see for example \cite{Giles:1977mp,Thorn:1979gu,Thorn:1997jy}),
but including interactions is frequently problematic.  The bit model
\cite{Verlinde:2002ig,Vaman:2002ka}
is intended as a step in the direction of constructing $H(J)$.

Perhaps, however, the best way to make $H(J)$ precise is simply to
read it off from the gauge theory!  Comparing matrix elements of
$\Delta - J$ in the gauge theory to matrix elements of~(\ref{eq:contact})
would let us read off the contact terms perturbatively
in the $\lambda' \sim 1/\mu^2$ expansion.

\section{Summary}

In these lectures we have learned that the essence of light-cone
string field theory is an interacting string Hamiltonian which satisfies
Properties 1 and 2 from Lecture 3.
Furthermore, we learned how to construct a Hamiltonian
which satsfies these properties, except possibly for measure
zero contact interactions for the superstring.
We also learned a little bit about the BMN limit of the
${\cal{N}} = 4$ SU($N$) Yang-Mills theory.
An obvious goal, which has been realized in a number of papers,
is to check whether the BMN limit of the gauge theory can
reproduce string amplitudes in the plane
wave background.
We now summarize the successses of this research program, and
end the lectures with a list of open questions.

\subsection{String Interactions in the BMN Correspondence}

Following closely the original construction of \cite{Green:hw}
in flat space, the three-string vertex for type IIB superstring field
theory in the plane-wave background has been constructed
\cite{Spradlin:2002ar,Spradlin:2002rv,Pankiewicz:2002tg}, including
the bosonic and fermionic Neumann coefficients, matrix elements of the
prefactor, and even explicit formulas for the Neumann coefficients
to all orders in $\lambda'$ \cite{He:2002zu}.
Matrix elements of the cubic interaction in string field theory have
been successfully matched to matrix elements
of $\Delta -J$ in the dual gauge theory after taking into account the relevant
basis transformation discussed above \cite{Gross:2002mh,
Pearson:2002zs,Gomis:2002wi,Roiban:2002xr}.
This program is highly developed and further aspects of this
approach have been studied in
\cite{Constable:2002hw,Kiem:2002xn,Huang:2002wf,Lee:2002rm,
Klebanov:2002mp,Huang:2002yt,Gursoy:2002yy,Schwarz:2002bc,
Pankiewicz:2002gs,Constable:2002vq,Lee:2002vz,
Janik:2002bd,
Kiem:2002pb,
Gursoy:2002fj,Gomis:2003kj,Bobkov:2003vg,Spradlin:2003bw,
Jr:2003zc,Freedman:2003bh,Gomis:2003kb,Georgiou:2003kt,
Pankiewicz:2003pg,Callan:2003xr,
Pankiewicz:2003ap}.

All of the successful checks of this correspondence have been restricted
to amplitudes where the number of impurities (in the gauge
theory language) is conserved.
For such amplitudes, all calculations so far indicate
that the gauge theory can indeed match the string theory prediction,
even though we are in a large 't Hooft coupling limit of the gauge
theory.  
The successful match works quite generally at first order in $\lambda'$
and $g_2$, and
at order $g_2^2$ (when intermediate impurity number violating
processes are omitted).
Although there is no obstacle in principle to pushing these checks
to higher order, the light-cone string field theory becomes
prohibitively complicated.

Since it is obvious that 
the repeated splitting and joining (i.e., $k$-trace operators
to $k{\pm}1$-trace operators) in the gauge theory provides, in the large
$J$ limit, a triangulation of the string theory moduli space (Property 1),
these successful tests of the BMN
correspondence amount to checking that the gauge theory
also knows about the correct measure on moduli space (Property 2),
at least in some limits of $\lambda'$ and $g_2$ parameter space.

\subsection{Some Open Questions and Puzzles}

We conclude our lectures with a partial list of open problems
and interesting directions for further research.

\bigskip
\noindent
{\bf Explore the Structure of the String Field  Theory.}
Can we say more about light-cone string field theory in the plane wave background?
In particular, is it possible to determine the measure factor
$f(p_1^+,p_2^+,p_3^+,\mu)$?
Is it possible to fully calculate a 4-particle interaction, or a 1-loop
mass renormalization?    What can be said about the contact terms in
the large $\mu$ limit?

\bigskip
\noindent
{\bf Does an $S$-matrix Exist in the Plane Wave?}
See references \cite{Dobashi:2002ar,Bak:2002ku,D'Appollonio:2003dr,
Cheung:2003ym,Yoneya:2003mu}.
Obviously, we have assumed in these lectures that the answer is yes.
The existence of an $S$-matrix has several
very interesting
consequences which should be possible to check purely within
the gauge theory.
For example, it implies that the one-loop mass renormalization of any $k$-particle
state should be equal to the sum of the one-loop mass renormalizations of
the individual particles.  This is because the existence of an $S$-matrix
presupposes that the particles can be well-separated from each other
(in the $x^-$ direction).
For the most trivial case, where one particle has 2 impurities, and the other
$k{-}1$ particles are all in the ground state, ${\rm Tr}[Z^J]$, this has indeed
been shown to be true (it is essentially due to the fact that `disconnected'
diagrams dominate over `connected' ones in the large $J$ limit---see
\cite{Gursoy:2002fj} for details).  Can this proof be generalized
to $k$ operators, each of which has more than zero (ideally, arbitrarily
many) impurities?

\bigskip

\noindent
{\bf Calculate the Gauge Theory Inner Product.}
The gauge theory inner product is defined
as the coefficient of the free ($g_{\rm YM} = 0$) two-point function.  
For example, in the simplest case of two vacuum operators
it has been shown that \cite{Kristjansen:2002bb,Constable:2002hw}
\begin{equation}
\langle {\rm Tr} [\overline{Z}^J] {\rm Tr} [Z^J ]\rangle = {\sinh(g_2/2) \over g_2/2}.
\end{equation}
For BMN operators with impurities, the inner product has been calculated only
to a couple of orders in $g_2$.
Since this is a free gauge theory calculation, it reduces to a simple
Gaussian matrix model, with graphs of genus $g$ contributing at order $g_2^{2 g}$.
It should be possible to prove (or disprove!) the conjectured formula that
the inner product is given in general by $e^{g_2 \Sigma}$, where $\Sigma$ is
the splitting-joining operator of the bit model.

\bigskip
\noindent
{\bf `Solve' the Quantum Mechanics of BMN Operators.}
Recent studies
have uncovered hints of integrable structures in
${\cal{N}}=4$ gauge theory.
One interesting question is whether there is any more hidden structure
in the quantum mechanical Hamiltonian we wrote down.  For example, one can check
(at least in the two-impurity sector) that the interaction $W$ in the string
Hamiltonian ${\widetilde{H}}$ commutes with $\Sigma$
\cite{Spradlin:2003bw}!  Are there any other
hidden symmetries which will allow one to make progress towards solving
this quantum mechanics?

\bigskip
\noindent
{\bf Do Higher-Point Functions Play any Role in the BMN Limit?}
In our discussion we limited our interest purely to two-point functions in
the gauge theory (although we did consider two-point functions of $k$-trace operators
with $l$-trace operators, but that contains only a tiny remnant of
the information encoded in the $k+l$-point function).
These two-point functions were interpreted (essentially) as $S$-matrix elements.
It is natural to wonder (as many papers have) what (if any) role is played
by the higher-point functions.

In the previous lecture we explained the correspondence between three-point
functions,  matrix elements of $|V\rangle$, and matrix
elements of $\Sigma$ at $\lambda' = 0$.  We avoided attempting to extend
this correspondence to $\lambda'> 0$ for the following reason.
In a conformal field theory, three-point functions of operators with well-defined
scaling dimension have simple three-point functions.  But BMN operators do
{\bf not} have well-defined scaling dimension when
$g_2>0$!  They are eigenstates of
the `free' Hamiltonian $H_0$,
but they are not eigenstates of $H = \Delta - J$.
So in general, the three-point function of BMN operators looks like
\begin{equation}
\label{eq:three}
\langle O_1(x_1) O_2(x_2) O_3(x_3)\rangle =
{\mbox{very complicated function of }}(x_1,x_2,x_3).
\end{equation}
We do not know how to recover the $x_1$, $x_2$, $x_3$ dependence of~(\ref{eq:three})
from the string theory side.

Another point to keep in mind is that higher-point functions of BMN operators
typically diverge in the large $J$ limit
\cite{Kristjansen:2002bb}.  An interesting alternative is
to consider $n$-point functions where 2 operators are BMN, and $n{-}2$
operators
have finite charge (for example, ${\rm Tr}[\phi Z]$).
Such an $n$-point function might be interpreted as an amplitude for propagation
from some initial state to some final state, with $n{-}2$
other operators inserted
on the worldsheet which perturb the spacetime away from the pure pp-wave
background.  The correspondence between spacetime perturbations and operator
insertions can be read off from the familiar dictionary of $AdS_5 \times S^5$.
Some calculations along these lines have been performed in
\cite{Mann:2003qp,us}.

\bigskip
\noindent
{\bf Define $H(J)$ Precisely and Provide a String Theory Construction
for It.}
Equivalently,
{\bf Can Continuum Light-Cone Superstring Field Theory be Honestly Discretized?}
Discretized string theories have been studied for a long time (in particular
by Thorn \cite{Giles:1977mp,Thorn:1979gu,Thorn:1997jy}),
but it seems somewhat problematic to discretize an interacting
type IIB string.  In particular, there are problems with the fermions.
There is the usual fermion doubling problem
(see \cite{Bellucci:2003qi,Danielsson:2003yc,Bellucci:2003hq})
that was
mentioned briefly in the lectures of H.~Verlinde.
There is also the basic question of how to implement the hermiticity
constraint~(\ref{eq:hermiticity}),
which in the continuum theory singles out the fermionic zero modes
from the non-zero modes.  However, a discretized string doesn't really have
any fermionic zero modes, there is just one fermionic oscillator on each
`site' along the string, and the natural `adjoint' just takes the adjoint
of each fermion at each site, in contrast with~(\ref{eq:hermiticity}).

\bigskip
\noindent
{\bf Which Quantities can be Calculated Perturbatively in the Gauge Theory?}
It is important to not forget that the BMN limit still involves taking
the 't Hooft coupling to infinity, so the weak coupling expansion is not
really valid.  However, it is empirically observed that some quantities, notably
the conformal dimensions of BMN operators at $g_2 = 0$:
\begin{equation}
\Delta - J = \sum_{n=-\infty}^\infty N_n \sqrt{1 + \lambda' J^2},
\end{equation}
may be calculated at small $\lambda$ (i.e., in gauge theory perturbation theory)
and finite $J$, and then extrapolated to $
\lambda, J \to \infty$ where magically they agree with the corresponding
string calculation!

These quantities are not BPS, so we had no right to expect this miracle to occur.
The basic question behind the BMN correspondence is simply this:  for which (if any!)
other quantities does this miracle occur?
For example, we commented that there have been successful comparisons of
off-shell matrix elements of the light-cone Hamiltonian, at one loop (order
$\lambda'$),  but we don't know if this miracle will continue to hold
for more complicated quantities.  This question is essentially the same
as:

\bigskip
\noindent
{\bf How is the Order of Limits Problem Resolved?}
The order of limits problem is that in the gauge theory, we want to
expand around $\lambda'  = 0$, which on the string theory corresponds
to $\mu = \infty$ and seems to be quite a singular limit.
In particular, several steps in the derivation of the light-cone string
vertex depended on the assertion that at small distances on the worldsheet,
the physics is essentially unchanged by the addition of the parameter $\mu$.
However this doesn't really make sense if $\mu$ is strictly infinity.

In particular, the prefactor of the superstring is a local operator insertion
at the point where the string splits.
If we view the gauge theory as a discretized version of the string theory
(with $J$ `bits'), how can the prefactor possibly be recovered when all 
calculations are necessarily done at finite $J$, where there is no notion
of `locality' on the worldsheet?

The successful checks of the BMN correspondence in the literature
indicate that this problem is somehow resolved, at least to leading
order in $\lambda'$.  It is not known whether this continues to hold
at higher order in $\lambda'$.
Certainly, nothing guarantees that the BMN correspondence has
to work perturbatively in both parameters $\lambda'$ and $g_2$.
Clearly, what we would like to do is to somehow calculate gauge theory
quantities at finite $\lambda$ and $J$, and then take the limit
$\lambda,J \to \infty$ holding $\lambda' = \lambda/J^2$ fixed.
This leads to the question:

\bigskip
\noindent
{\bf Can We Turn on Finite $\lambda$ in the Bit Model?}
One way that it might be possible to make sense of
working at finite $\lambda$ in the gauge theory is to
work at finite $\lambda$
in the bit model \cite{Vaman:2002ka},
with the hope that the bit model successfully encapsulates
all of the relevant gauge theory degrees of freedom.

Let $a_x$, $a_x^\dagger$ denote the usual operators living on site $x$ on the string
and satisfying $[a_x, a_y^\dagger] = \delta_{xy}$.
At $\lambda = 0$, these operators are just raising and lowering operators
of the Hamiltonian.  That is, creating an excitation at some position $x$
on the string precisely increases the energy by one unit.

However at $\lambda \ne 0$ it is no longer the case that the Hamiltonian
is diagonal in this `local' basis.  Instead, the Hamiltonian is diagonal
in the Fourier basis, and the eigenmodes are not those which are located
at some point $x$ on the string, but rather are those with well-defined
momentum $n$ around the string.

One can turn on finite $\lambda$ in the bit model by doing a Bogolyubov
transformation; the operator $a_x$ will now be expressed in terms of
both raising and lowering operators $a_n$ and $a_n^\dagger$.  The cubic
vertex, which is trivial at $\lambda = 0$ because it is just the
delta-function overlap matching excitations pointwise in $\sigma$, needs
to be reexpressed in terms of the true eigenmodes of the $\lambda>0$
Hamiltonian.

\bigskip
\noindent
{\bf Is there Another Tool?}
Rather than attempting to compare gauge theory calculations
to the known but complicated
\cite{Green:hw} formulation of string field theory,
it might also be possible to  derive a string field theory
directly from the gauge theory---i.e., to show that
the large $J$ limit of gauge theory correlation functions
gives a correct measure on moduli space.
Some work which is
more along these lines includes
\cite{Berenstein:2002sa,deMelloKoch:2002nq,Dobashi:2002ar,
Yoneya:2003mu,Dijkgraaf:2003nw}.

\section*{Acknowledgments}

It is a pleasure to thank Y.~He, I.~Klebanov, J.~Pearson,
R.~Roiban, J.~Schwarz,
D.~Vaman and H.~Verlinde for collaboration on material presented
in these lectures.
M.S.~would like to thank the organizers and staff of the
Trieste Spring School on Superstring Theory and Related Topics
for hosting a very stimulating workshop.
This work has been supported in part
by the DOE under grant DE-FG02-91ER40671,
and by
the National Science Foundation under Grant No.~PHY99-07949.
Any opinions, findings, and conclusions or recommendations expressed
in this material are those of the authors and do not necessarily
reflect the views of the National Science Foundation.

%%%%%%%%%%%%%%%%%%%%%%%%%%%%%%%%%%%%%%%%%%%%%%%%%%%%%%%%%%%%%%%%%%%%%%%%%%%
%%%            Appendices (if any) start here                             %
%%%%%%%%%%%%%%%%%%%%%%%%%%%%%%%%%%%%%%%%%%%%%%%%%%%%%%%%%%%%%%%%%%%%%%%%%%%
%%% IF YOU DO NOT HAVE ANY APPENDIX, 
%%% REMOVE EVERYTHING TILL "End of Appendixes"

%%%           End of Appendixes						  %
%%%%%%%%%%%%%%%%%%%%%%%%%%%%%%%%%%%%%%%%%%%%%%%%%%%%%%%%%%%%%%%%%%%%%%%%%%%

%%%%%%%%%%%%%%%%%%%%%%%%%%%%%%%%%%%%%%%%%%%%%%%%%%%%%%%%%%%%%%%%%%%%%%%%%%%
%%%           References starts here                                      %
%%%%%%%%%%%%%%%%%%%%%%%%%%%%%%%%%%%%%%%%%%%%%%%%%%%%%%%%%%%%%%%%%%%%%%%%%%%


\newpage
\addcontentsline{toc}{section}{References}
\begin{thebibliography}{999}

\bibitem{Mandelstam:jk}
S.~Mandelstam,
``Interacting String Picture Of Dual Resonance Models,''
Nucl.\ Phys.\ B {\bf 64} (1973) 205.
%%CITATION = NUPHA,B64,205;%%

\bibitem{Mandelstam:hk}
S.~Mandelstam,
``Interacting String Picture Of The Neveu-Schwarz-Ramond Model,''
Nucl.\ Phys.\ B {\bf 69} (1974) 77.
%%CITATION = NUPHA,B69,77;%%

\bibitem{Cremmer:1974jq}
E.~Cremmer and .~L.~Gervais,
``Combining And Splitting Relativistic Strings,''
Nucl.\ Phys.\ B {\bf 76} (1974) 209.
%%CITATION = NUPHA,B76,209;%%

\bibitem{Cremmer:1974ej}
E.~Cremmer and .~L.~Gervais,
``Infinite Component Field Theory Of Interacting Relativistic Strings
And Dual Theory,''
Nucl.\ Phys.\ B {\bf 90} (1975) 410.
%%CITATION = NUPHA,B90,410;%%

\bibitem{Giles:1977mp}
R.~Giles and C.~B.~Thorn,
``A Lattice Approach To String Theory,''
Phys.\ Rev.\ D {\bf 16} (1977) 366.
%%CITATION = PHRVA,D16,366;%%

\bibitem{Thorn:1979gu}
C.~B.~Thorn,
``A Fock Space Description Of The $1/N_C$ Expansion Of Quantum Chromodynamics,''
Phys.\ Rev.\ D {\bf 20} (1979) 1435.
%%CITATION = PHRVA,D20,1435;%%

\bibitem{Green:1982tk}
M.~B.~Green and J.~H.~Schwarz,
``Extended Supergravity In Ten-Dimensions,''
Phys.\ Lett.\ B {\bf 122} (1983) 143.
%%CITATION = PHLTA,B122,143;%%

\bibitem{Green:1982tc}
M.~B.~Green and J.~H.~Schwarz,
``Superstring Interactions,''
Nucl.\ Phys.\ B {\bf 218} (1983) 43.
%%CITATION = NUPHA,B218,43;%%

\bibitem{Green:hw}
M.~B.~Green, J.~H.~Schwarz and L.~Brink,
``Superfield Theory Of Type II Superstrings,''
Nucl.\ Phys.\ B {\bf 219} (1983) 437.
%%CITATION = NUPHA,B219,437;%%

\bibitem{Green:fu}
M.~B.~Green and J.~H.~Schwarz,
``Superstring Field Theory,''
Nucl.\ Phys.\ B {\bf 243} (1984) 475.
%%CITATION = NUPHA,B243,475;%%

\bibitem{Schwarz}
J.~H.~Schwarz (Editor),
``Superstrings: The First 15 Years of Superstring Theory, Volume 2''
World Scientific (1985).

\bibitem{GSW}
M.~B.~Green, J.~H.~Schwarz and E.~Witten,
``Superstring Theory, Vol. 2: Loop Amplitudes, Anomalies And Phenomenology,''
Cambridge University Press (1987).

\bibitem{Giddings:1986rf}
S.~B.~Giddings and S.~A.~Wolpert,
``A Triangulation Of Moduli Space From Light Cone String Theory,''
Commun.\ Math.\ Phys.\  {\bf 109} (1987) 177.
%%CITATION = CMPHA,109,177;%%

\bibitem{Greensite:1986gv}
J.~Greensite and F.~R.~Klinkhamer,
``New Interactions For Superstrings,''
Nucl.\ Phys.\ B {\bf 281} (1987) 269.
%%CITATION = NUPHA,B281,269;%%

\bibitem{D'Hoker:1987pr}
E.~D'Hoker and S.~B.~Giddings,
``Unitarity Of The Closed Bosonic Polyakov String,''
Nucl.\ Phys.\ B {\bf 291} (1987) 90.
%%CITATION = NUPHA,B291,90;%% 

\bibitem{Greensite:1987sm}   
J.~Greensite and F.~R.~Klinkhamer,
``Contact Interactions In Closed Superstring Field Theory,''
Nucl.\ Phys.\ B {\bf 291} (1987) 557.   
%%CITATION = NUPHA,B291,557;%%

\bibitem{Green:1987qu}
M.~B.~Green and N.~Seiberg,
``Contact Interactions In Superstring Theory,''
Nucl.\ Phys.\ B {\bf 299} (1988) 559.
%%CITATION = NUPHA,B299,559;%%

\bibitem{Greensite:1987hm}
J.~Greensite and F.~R.~Klinkhamer,
``Superstring Amplitudes And Contact Interactions,''
Nucl.\ Phys.\ B {\bf 304} (1988) 108.
%%CITATION = NUPHA,B304,108;%%

\bibitem{Thorn:1997jy}
C.~B.~Thorn,
``Supersymmetric quantum mechanics for string-bits,''
Phys.\ Rev.\ D {\bf 56} (1997) 6619
[arXiv:hep-th/9707048].
%%CITATION = HEP-TH 9707048;%%

\bibitem{Maldacena:1997re}
J.~M.~Maldacena,
``The large N limit of superconformal field theories and supergravity,''
Adv.\ Theor.\ Math.\ Phys.\  {\bf 2}, 231 (1998)
[Int.\ J.\ Theor.\ Phys.\  {\bf 38}, 1113 (1999)]
[arXiv:hep-th/9711200].
%%CITATION = HEP-TH 9711200;%%

\bibitem{Metsaev:1998it}
R.~R.~Metsaev and A.~A.~Tseytlin,
``Type IIB superstring action in AdS${}_5 \times$ S${}^5$ background,''
Nucl.\ Phys.\ B {\bf 533} (1998) 109
[arXiv:hep-th/9805028].
%%CITATION = HEP-TH 9805028;%%

\bibitem{Kallosh:1998zx}
R.~Kallosh, J.~Rahmfeld and A.~Rajaraman,
``Near horizon superspace,''
JHEP {\bf 9809} (1998) 002
[arXiv:hep-th/9805217].
%%CITATION = HEP-TH 9805217;%%

\bibitem{Aharony:1999ti}
O.~Aharony, S.~S.~Gubser, J.~M.~Maldacena, H.~Ooguri and Y.~Oz,
``Large N field theories, string theory and gravity,''
Phys.\ Rept.\  {\bf 323} (2000) 183
[arXiv:hep-th/9905111].
%%CITATION = HEP-TH 9905111;%%

\bibitem{Roiban:2000yy}
R.~Roiban and W.~Siegel,
``Superstrings on $AdS_5 \times S^5$ supertwistor space,''
JHEP {\bf 0011} (2000) 024
[arXiv:hep-th/0010104].
%%CITATION = HEP-TH 0010104;%%

\bibitem{Blau:2001ne}
M.~Blau, J.~Figueroa-O'Farrill, C.~Hull and G.~Papadopoulos,
``A new maximally supersymmetric background of IIB superstring theory,''
JHEP {\bf 0201} (2002) 047
[arXiv:hep-th/0110242].
%%CITATION = HEP-TH 0110242;%%

\bibitem{Metsaev:2001bj}
R.~R.~Metsaev,
``Type IIB Green-Schwarz superstring in plane wave Ramond-Ramond  background,''
Nucl.\ Phys.\ B {\bf 625} (2002) 70
[arXiv:hep-th/0112044].
%%CITATION = HEP-TH 0112044;%%

\bibitem{Blau:2002dy}
M.~Blau, J.~Figueroa-O'Farrill, C.~Hull and G.~Papadopoulos,
``Penrose limits and maximal supersymmetry,''
Class.\ Quant.\ Grav.\  {\bf 19} (2002) L87
[arXiv:hep-th/0201081].
%%CITATION = HEP-TH 0201081;%%

\bibitem{D'Hoker:2002aw}
E.~D'Hoker and D.~Z.~Freedman,
``Supersymmetric gauge theories and the AdS/CFT correspondence,''
arXiv:hep-th/0201253.
%%CITATION = HEP-TH 0201253;%%

\bibitem{Berenstein:2002jq}
D.~Berenstein, J.~M.~Maldacena and H.~Nastase,
``Strings in flat space and pp waves from ${\cal{N}} = 4$ super Yang Mills,''
JHEP {\bf 0204} (2002) 013
[arXiv:hep-th/0202021].
%%CITATION = HEP-TH 0202021;%%

\bibitem{Metsaev:2002re}
R.~R.~Metsaev and A.~A.~Tseytlin,
``Exactly solvable model of superstring in plane wave Ramond-Ramond
 background,''
Phys.\ Rev.\ D {\bf 65} (2002) 126004
[arXiv:hep-th/0202109].
%%CITATION = HEP-TH 0202109;%%

\bibitem{Blau:2002mw}
M.~Blau, J.~Figueroa-O'Farrill and G.~Papadopoulos,
``Penrose limits, supergravity and brane dynamics,''
Class.\ Quant.\ Grav.\  {\bf 19} (2002) 4753
[arXiv:hep-th/0202111].
%%CITATION = HEP-TH 0202111;%%

\bibitem{Russo:2002rq}
J.~G.~Russo and A.~A.~Tseytlin,
``On solvable models of type IIB superstring in NS-NS and R-R plane wave
backgrounds,''
JHEP {\bf 0204} (2002) 021
[arXiv:hep-th/0202179].
%%CITATION = HEP-TH 0202179;%%

\bibitem{Hatsuda:2002xp}
M.~Hatsuda, K.~Kamimura and M.~Sakaguchi,
``From super-$AdS_5 \times S^5$ algebra to super-pp-wave algebra,''
Nucl.\ Phys.\ B {\bf 632} (2002) 114
[arXiv:hep-th/0202190].
%%CITATION = HEP-TH 0202190;%%

\bibitem{Michelson:2002wa}
J.~Michelson,
``(Twisted) toroidal compactification of pp-waves,''
Phys.\ Rev.\ D {\bf 66} (2002) 066002
[arXiv:hep-th/0203140].
%%CITATION = HEP-TH 0203140;%%

\bibitem{Das:2002cw}
S.~R.~Das, C.~Gomez and S.~J.~Rey,
``Penrose limit, spontaneous symmetry breaking and holography in pp-wave
background,''
Phys.\ Rev.\ D {\bf 66} (2002) 046002
[arXiv:hep-th/0203164].
%%CITATION = HEP-TH 0203164;%%

\bibitem{Berkovits:2002zv}
N.~Berkovits,
``Conformal field theory for the superstring in a Ramond-Ramond plane
wave background,''
JHEP {\bf 0204} (2002) 037
[arXiv:hep-th/0203248].
%%CITATION = HEP-TH 0203248;%%

\bibitem{Kiritsis:2002kz}
E.~Kiritsis and B.~Pioline,
``Strings in homogeneous gravitational waves and null holography,''
JHEP {\bf 0208} (2002) 048
[arXiv:hep-th/0204004].
%%CITATION = HEP-TH 0204004;%%

\bibitem{Leigh:2002pt}
R.~G.~Leigh, K.~Okuyama and M.~Rozali,
``PP-waves and holography,''
Phys.\ Rev.\ D {\bf 66} (2002) 046004
[arXiv:hep-th/0204026].
%%CITATION = HEP-TH 0204026;%%

\bibitem{Spradlin:2002ar}
M.~Spradlin and A.~Volovich,
``Superstring interactions in a pp-wave background,''
Phys.\ Rev.\ D {\bf 66} (2002) 086004
[arXiv:hep-th/0204146].
%%CITATION = HEP-TH 0204146;%%

\bibitem{Kristjansen:2002bb}
C.~Kristjansen, J.~Plefka, G.~W.~Semenoff and M.~Staudacher,
``A new double-scaling limit of
${\cal{N}} = 4$ super Yang-Mills theory and PP-wave
strings,''
Nucl.\ Phys.\ B {\bf 643} (2002) 3
[arXiv:hep-th/0205033].
%%CITATION = HEP-TH 0205033;%%

\bibitem{Berenstein:2002sa}
D.~Berenstein and H.~Nastase,
``On lightcone string field theory from super Yang-Mills and holography,''
arXiv:hep-th/0205048.
%%CITATION = HEP-TH 0205048;%%

\bibitem{Gross:2002su}
D.~J.~Gross, A.~Mikhailov and R.~Roiban,
``Operators with large R charge in N = 4 Yang-Mills theory,''
Annals Phys.\  {\bf 301} (2002) 31
[arXiv:hep-th/0205066].
%%CITATION = HEP-TH 0205066;%%

\bibitem{Constable:2002hw}
N.~R.~Constable, D.~Z.~Freedman, M.~Headrick, S.~Minwalla, L.~Motl,
A.~Postnikov and W.~Skiba,
``PP-wave string interactions from perturbative Yang-Mills theory,''
JHEP {\bf 0207} (2002) 017
[arXiv:hep-th/0205089].
%%CITATION = HEP-TH 0205089;%%

\bibitem{Gopakumar:2002dq}
R.~Gopakumar,
``String interactions in PP-waves,''
Phys.\ Rev.\ Lett.\  {\bf 89} (2002) 171601
[arXiv:hep-th/0205174].
%%CITATION = HEP-TH 0205174;%%

\bibitem{Kiem:2002xn}
Y.~Kiem, Y.~Kim, S.~Lee and J.~Park,
``pp-wave / Yang-Mills correspondence: An explicit check,''
Nucl.\ Phys.\ B {\bf 642} (2002) 389
[arXiv:hep-th/0205279].
%%CITATION = HEP-TH 0205279;%%

\bibitem{Huang:2002wf}
M.~Huang,
``Three point functions of
${\cal{N}} = 4$
super Yang Mills from light cone string  field theory in pp-wave,''
Phys.\ Lett.\ B {\bf 542} (2002) 255
[arXiv:hep-th/0205311].
%%CITATION = HEP-TH 0205311;%%

\bibitem{Chu:2002pd}
C.~S.~Chu, V.~V.~Khoze and G.~Travaglini,
``Three-point functions in ${\cal{N}} = 4$ Yang-Mills theory and pp-waves,''
JHEP {\bf 0206} (2002) 011
[arXiv:hep-th/0206005].
%%CITATION = HEP-TH 0206005;%%

\bibitem{Verlinde:2002ig}
H.~Verlinde,
``Bits, matrices and 1/N,''
arXiv:hep-th/0206059.
%%CITATION = HEP-TH 0206059;%%

\bibitem{Lee:2002rm}
P.~Lee, S.~Moriyama and J.~Park,
``Cubic interactions in pp-wave light cone string field theory,''
Phys.\ Rev.\ D {\bf 66} (2002) 085021
[arXiv:hep-th/0206065].
%%CITATION = HEP-TH 0206065;%%

\bibitem{Spradlin:2002rv}
M.~Spradlin and A.~Volovich,
``Superstring interactions in a pp-wave background. II,''
JHEP {\bf 0301} (2003) 036
[arXiv:hep-th/0206073].
%%CITATION = HEP-TH 0206073;%%

\bibitem{Santambrogio:2002sb}
A.~Santambrogio and D.~Zanon,
``Exact anomalous dimensions of ${\cal{N}} = 4$
Yang-Mills operators with large $R$  charge,''
Phys.\ Lett.\ B {\bf 545} (2002) 425
[arXiv:hep-th/0206079].
%%CITATION = HEP-TH 0206079;%%

\bibitem{Chu:2002qj}
C.~S.~Chu, V.~V.~Khoze and G.~Travaglini,
``pp-wave string interactions from $n$-point correlators of BMN operators,''
JHEP {\bf 0209} (2002) 054
[arXiv:hep-th/0206167].
%%CITATION = HEP-TH 0206167;%%

\bibitem{Klebanov:2002mp}
I.~R.~Klebanov, M.~Spradlin and A.~Volovich,
``New effects in gauge theory from pp-wave superstrings,''
Phys.\ Lett.\ B {\bf 548} (2002) 111
[arXiv:hep-th/0206221].
%%CITATION = HEP-TH 0206221;%%

\bibitem{Huang:2002yt}
M.~Huang,
``String interactions in pp-wave from ${\cal{N}} = 4$ super Yang Mills,''
Phys.\ Rev.\ D {\bf 66} (2002) 105002
[arXiv:hep-th/0206248].
%%CITATION = HEP-TH 0206248;%%

\bibitem{Parnachev:2002kk}
A.~Parnachev and A.~V.~Ryzhov,
``Strings in the near plane wave background and AdS/CFT,''
JHEP {\bf 0210} (2002) 066
[arXiv:hep-th/0208010].
%%CITATION = HEP-TH 0208010;%%

\bibitem{Gursoy:2002yy}
U.~Gursoy,
``Vector operators in the BMN correspondence,''
arXiv:hep-th/0208041.
%%CITATION = HEP-TH 0208041;%%

\bibitem{Berkovits:2002vn}
N.~Berkovits and J.~Maldacena,
``N = 2 superconformal description of superstring in Ramond-Ramond plane
wave backgrounds,''
JHEP {\bf 0210} (2002) 059
[arXiv:hep-th/0208092].
%%CITATION = HEP-TH 0208092;%%

\bibitem{Chu:2002eu}
C.~S.~Chu, V.~V.~Khoze, M.~Petrini, R.~Russo and A.~Tanzini,
``A note on string interaction on the pp-wave background,''
arXiv:hep-th/0208148.
%%CITATION = HEP-TH 0208148;%%

\bibitem{Beisert:2002bb}
N.~Beisert, C.~Kristjansen, J.~Plefka, G.~W.~Semenoff and M.~Staudacher,
``BMN correlators and operator mixing in N = 4 super Yang-Mills theory,''
Nucl.\ Phys.\ B {\bf 650} (2003) 125
[arXiv:hep-th/0208178].
%%CITATION = HEP-TH 0208178;%%

\bibitem{Schwarz:2002bc}
J.~H.~Schwarz,
``Comments on superstring interactions in a plane wave background,''
JHEP {\bf 0209} (2002) 058
[arXiv:hep-th/0208179].
%%CITATION = HEP-TH 0208179;%%

\bibitem{Pankiewicz:2002gs}
A.~Pankiewicz,
``More comments on superstring interactions in the pp-wave background,''
JHEP {\bf 0209} (2002) 056
[arXiv:hep-th/0208209].
%%CITATION = HEP-TH 0208209;%%

\bibitem{Gross:2002mh}
D.~J.~Gross, A.~Mikhailov and R.~Roiban,
``A calculation of the plane wave string Hamiltonian from N = 4
super-Yang-Mills theory,''
arXiv:hep-th/0208231.
%%CITATION = HEP-TH 0208231;%%

\bibitem{Zhou:2002mi}
J.-G.~Zhou,
``pp-wave string interactions from string bit model,''
Phys.\ Rev.\ D {\bf 67} (2003) 026010
[arXiv:hep-th/0208232].
%%CITATION = HEP-TH 0208232;%%

\bibitem{Constable:2002vq}
N.~R.~Constable, D.~Z.~Freedman, M.~Headrick and S.~Minwalla,
``Operator mixing and the BMN correspondence,''
JHEP {\bf 0210} (2002) 068
[arXiv:hep-th/0209002].
%%CITATION = HEP-TH 0209002;%%

\bibitem{Lee:2002vz}
P.~Lee, S.~Moriyama and J.~Park,
``A note on cubic interactions in pp-wave light cone string field theory,''
Phys.\ Rev.\ D {\bf 67} (2003) 086001
[arXiv:hep-th/0209011].
%%CITATION = HEP-TH 0209011;%%

\bibitem{deMelloKoch:2002nq}
R.~de Mello Koch, A.~Jevicki and J.~P.~Rodrigues,
``Collective string field theory of matrix models in the BMN limit,''
arXiv:hep-th/0209155.
%%CITATION = HEP-TH 0209155;%%

\bibitem{Vaman:2002ka}
D.~Vaman and H.~Verlinde,
``Bit strings from N = 4 gauge theory,''
arXiv:hep-th/0209215.
%%CITATION = HEP-TH 0209215;%%

\bibitem{Eynard:2002df}
B.~Eynard and C.~Kristjansen,
``BMN correlators by loop equations,''
JHEP {\bf 0210} (2002) 027
[arXiv:hep-th/0209244].
%%CITATION = HEP-TH 0209244;%%

\bibitem{Dobashi:2002ar}
S.~Dobashi, H.~Shimada and T.~Yoneya,
``Holographic reformulation of string theory on AdS${}_5 \times$ S${}^5$
background in the PP-wave limit,''
Nucl.\ Phys.\ B {\bf 665} (2003) 94
[arXiv:hep-th/0209251].
%%CITATION = HEP-TH 0209251;%%

\bibitem{Janik:2002bd}
R.~A.~Janik,
``BMN operators and string field theory,''
Phys.\ Lett.\ B {\bf 549} (2002) 237
[arXiv:hep-th/0209263].
%%CITATION = HEP-TH 0209263;%%

\bibitem{Pearson:2002zs}
J.~Pearson, M.~Spradlin, D.~Vaman, H.~Verlinde and A.~Volovich,
``Tracing the string: BMN correspondence at finite $J^2/N$,''
arXiv:hep-th/0210102.
%%CITATION = HEP-TH 0210102;%%

\bibitem{Gomis:2002wi}
J.~Gomis, S.~Moriyama and J.~Park,
``SYM description of SFT Hamiltonian in a pp-wave background,''
arXiv:hep-th/210153.
%%CITATION = HEP-TH 0210153;%%

\bibitem{Pankiewicz:2002tg}
A.~Pankiewicz and B.~Stefa\'nski,
``pp-wave light-cone superstring field theory,''
Nucl.\ Phys.\ B {\bf 657} (2003) 79
[arXiv:hep-th/0210246].
%%CITATION = HEP-TH 0210246;%%

\bibitem{Beisert:2002tn}
N.~Beisert,
``BMN operators and superconformal symmetry,''
arXiv:hep-th/0211032.
%%CITATION = HEP-TH 0211032;%%

\bibitem{Bak:2002ku}
D.~Bak and M.~M.~Sheikh-Jabbari,
``Strong evidence in favor of the existence of
S-matrix for strings in  plane waves,''
JHEP {\bf 0302} (2003) 019
[arXiv:hep-th/0211073].
%%CITATION = HEP-TH 0211073;%%

\bibitem{Chu:2002wj}
C.~S.~Chu, M.~Petrini, R.~Russo and A.~Tanzini,
``String interactions and discrete symmetries of the pp-wave background,''
Class.\ Quant.\ Grav.\  {\bf 20} (2003) S457
[arXiv:hep-th/0211188].
%%CITATION = HEP-TH 0211188;%%

\bibitem{He:2002zu}
Y.~H.~He, J.~H.~Schwarz, M.~Spradlin and A.~Volovich,
``Explicit formulas for Neumann coefficients in the plane wave geometry,''
arXiv:hep-th/0211198.
%%CITATION = HEP-TH 0211198;%%

\bibitem{Kiem:2002pb}
Y.~Kiem, Y.~Kim, J.~Park and C.~Ryou,
``Chiral primary cubic interactions from pp-wave supergravity,''
JHEP {\bf 0301} (2003) 026
[arXiv:hep-th/0211217].
%%CITATION = HEP-TH 0211217;%%

\bibitem{Roiban:2002xr}
R.~Roiban, M.~Spradlin and A.~Volovich,
``On light-cone SFT contact terms in a plane wave,''
arXiv:hep-th/0211220.
%%CITATION = HEP-TH 0211220;%%

\bibitem{Gursoy:2002fj}
U.~Gursoy,
``Predictions for pp-wave string amplitudes from perturbative SYM,''
arXiv:hep-th/0212118.
%%CITATION = HEP-TH 0212118;%%

\bibitem{Minahan:2002ve}
J.~A.~Minahan and K.~Zarembo,
``The Bethe-ansatz for ${\cal{N}} = 4$ super Yang-Mills,''
JHEP {\bf 0303} (2003) 013
[arXiv:hep-th/0212208].
%%CITATION = HEP-TH 0212208;%%

\bibitem{Beisert:2002ff}
N.~Beisert, C.~Kristjansen, J.~Plefka and M.~Staudacher,
``BMN gauge theory as a quantum mechanical system,''
Phys.\ Lett.\ B {\bf 558} (2003) 229
[arXiv:hep-th/0212269].
%%CITATION = HEP-TH 0212269;%%

\bibitem{Chu:2003qd}
C.~S.~Chu and V.~V.~Khoze,
``Correspondence between the 3-point BMN correlators
and the 3-string  vertex on the pp-wave,''
JHEP {\bf 0304} (2003) 014
[arXiv:hep-th/0301036].
%%CITATION = HEP-TH 0301036;%%

\bibitem{Gomis:2003kj}
J.~Gomis, S.~Moriyama and J.~Park,
``SYM description of pp-wave string interactions:
Singlet sector and  arbitrary impurities,''
Nucl.\ Phys.\ B {\bf 665} (2003) 49
[arXiv:hep-th/0301250].
%%CITATION = HEP-TH 0301250;%%

\bibitem{Bellucci:2003qi}
S.~Bellucci and C.~Sochichiu,
``Fermion doubling and BMN correspondence,''
Phys.\ Lett.\ B {\bf 564} (2003) 115
[arXiv:hep-th/0302104].
%%CITATION = HEP-TH 0302104;%%

\bibitem{Bobkov:2003vg}
K.~Bobkov,
``Graviton-scalar interaction in the pp-wave background,''
Mod.\ Phys.\ Lett.\ A {\bf 18} (2003) 1441
[arXiv:hep-th/0303007].
%%CITATION = HEP-TH 0303007;%%

\bibitem{Beisert:2003tq}
N.~Beisert, C.~Kristjansen and M.~Staudacher,
``The dilatation operator of ${\cal{N}} = 4$ super Yang-Mills theory,''
arXiv:hep-th/0303060.
%%CITATION = HEP-TH 0303060;%%

\bibitem{Chu:2003ji}
C.~S.~Chu, V.~V.~Khoze and G.~Travaglini,
``BMN operators with vector impurities, $Z_2$ symmetry and pp-waves,''
JHEP {\bf 0306} (2003) 050
[arXiv:hep-th/0303107].
%%CITATION = HEP-TH 0303107;%%

\bibitem{Spradlin:2003bw}
M.~Spradlin and A.~Volovich,
``Note on plane wave quantum mechanics,''
arXiv:hep-th/0303220.
%%CITATION = HEP-TH 0303220;%%

\bibitem{DiVecchia:2003yp}
P.~Di Vecchia, J.~L.~Petersen, M.~Petrini, R.~Russo and A.~Tanzini,
``The 3-string vertex and the AdS/CFT duality in the pp-wave limit,''
arXiv:hep-th/0304025.
%%CITATION = HEP-TH 0304025;%%

\bibitem{Jr:2003zc}
B.~Stefa\'nski~Jr.,
``Open string plane-wave light-cone superstring field theory,''
arXiv:hep-th/0304114.
%%CITATION = HEP-TH 0304114;%%

\bibitem{Yoneya:2003mu}
T.~Yoneya,
``What is holography in the plane-wave limit of $AdS_5$/SYM${}_4$
correspondence?,''
arXiv:hep-th/0304183.
%%CITATION = HEP-TH 0304183;%%

\bibitem{Chu:2003nc}
C.~S.~Chu and K.~Kyritsis,
``The string light cone in the pp-wave background,''
Phys.\ Lett.\ B {\bf 566} (2003) 240
[arXiv:hep-th/0304191].
%%CITATION = HEP-TH 0304191;%%

\bibitem{Freedman:2003bh}
D.~Z.~Freedman and U.~Gursoy,
``Instability and degeneracy in the BMN correspondence,''
JHEP {\bf 0308} (2003) 027
[arXiv:hep-th/0305016].
%%CITATION = HEP-TH 0305016;%%

\bibitem{D'Appollonio:2003dr}
G.~D'Appollonio and E.~Kiritsis,
``String interactions in gravitational wave backgrounds,''
arXiv:hep-th/0305081.
%%CITATION = HEP-TH 0305081;%%

\bibitem{Bena:2003wd}
I.~Bena, J.~Polchinski and R.~Roiban,
``Hidden symmetries of the $AdS_5 \times S^5$ superstring,''
arXiv:hep-th/0305116.
%%CITATION = HEP-TH 0305116;%%

\bibitem{Mann:2003qp}
N.~Mann and J.~Polchinski,
``AdS holography in the Penrose limit,''
arXiv:hep-th/0305230.
%%CITATION = HEP-TH 0305230;%%

\bibitem{us}
J.~Maldacena, M.~Spradlin and A.~Volovich, unpublished.

\bibitem{Gomis:2003kb}
J.~Gomis, S.~Moriyama and J.~Park,
``Open + closed string field theory from gauge fields,''
arXiv:hep-th/0305264.
%%CITATION = HEP-TH 0305264;%%

\bibitem{Danielsson:2003yc}
U.~Danielsson, F.~Kristiansson, M.~Lubcke and K.~Zarembo,
``String bits without doubling,''
arXiv:hep-th/0306147.
%%CITATION = HEP-TH 0306147;%%

\bibitem{Georgiou:2003kt}
G.~Georgiou, V.~V.~Khoze and G.~Travaglini,
``New tests of the pp-wave correspondence,''
arXiv:hep-th/0306234.
%%CITATION = HEP-TH 0306234;%%

\bibitem{Beisert:2003jj}
N.~Beisert,
``The complete one-loop dilatation operator of ${\cal{N}} = 4$ super Yang-Mills
theory,''
arXiv:hep-th/0307015.
%%CITATION = HEP-TH 0307015;%%

\bibitem{Pankiewicz:2003pg}
A.~Pankiewicz,
``Strings in plane wave backgrounds,''
arXiv:hep-th/0307027.
%%CITATION = HEP-TH 0307027;%%

\bibitem{Callan:2003xr}
C.~G.~Callan, H.~K.~Lee, T.~McLoughlin, J.~H.~Schwarz, I.~Swanson and X.~Wu,
``Quantizing string theory in $AdS_5 \times S^5$: Beyond the pp-wave,''
arXiv:hep-th/0307032.
%%CITATION = HEP-TH 0307032;%%

\bibitem{Plefka:2003nb}
J.~C.~Plefka,
``Lectures on the plane-wave string/gauge theory duality,''
arXiv:hep-th/0307101.
%%CITATION = HEP-TH 0307101;%%

\bibitem{Bellucci:2003hq}
S.~Bellucci and C.~Sochichiu,
``Can string bits be supersymmetric?,''
Phys.\ Lett.\ B {\bf 571} (2003) 92
[arXiv:hep-th/0307253].
%%CITATION = HEP-TH 0307253;%%

\bibitem{Pankiewicz:2003ap}
A.~Pankiewicz and B.~Stefa\'nski,
``On the uniqueness of plane-wave string field theory,''
arXiv:hep-th/0308062.
%%CITATION = HEP-TH 0308062;%%

\bibitem{Dolan:2003uh}
L.~Dolan, C.~R.~Nappi and E.~Witten,
``A relation between approaches to integrability in superconformal Yang-Mills
theory,''
arXiv:hep-th/0308089.
%%CITATION = HEP-TH 0308089;%%

\bibitem{Cheung:2003ym}
Y.~K.~Cheung, L.~Freidel and K.~Savvidy,
``Strings in gravimagnetic fields,''
arXiv:hep-th/0309005.
%%CITATION = HEP-TH 0309005;%%

\bibitem{Dijkgraaf:2003nw}
R.~Dijkgraaf and L.~Motl,
``Matrix string theory, contact terms, and superstring field theory,''
arXiv:hep-th/0309238.
%%CITATION = HEP-TH 0309238;%%

\end{thebibliography}
\end{document}